\documentclass[onecolumn]{pasj01}
\Received{$\langle$reception date$\rangle$}
\Accepted{$\langle$acception date$\rangle$}
\Published{$\langle$publication date$\rangle$}
\usepackage{ulem,slashbox,color,colortbl}
\pdfoutput=1
\begin{document}
\title{Extended Optical/NIR Observations of Type Iax Supernova 2014dt: Possible Signatures of a Bound Remnant \protect\thanks{
This work is based on data collected from the Subaru telescope, 
operated by the National Astronomical Observatory of Japan (NAOJ); 
the Kanata 1.5 m telescope, operated by Hiroshima University; 
and 51 cm telescope, operated by Osaka Kyoiku University.}}
%%%\author{Miho \textsc{Kawabata},\altaffilmark{1}}
%%%\altaffiltext{1}{Department of Physical Science, Hirosima University, Kagamiyama, Higashi-Hiroshima, Hiroshima 739-8526}
%%%\email{mkawabata@hep01.hepl.hiroshima-u.ac.jp}
\author{Miho \textsc{Kawabata}\altaffilmark{1}, 
Koji S. \textsc{Kawabata}\altaffilmark{2,1,3}, 
Keiich \textsc{Maeda}\altaffilmark{4,5}, 
Masayuki \textsc{Yamanaka}\altaffilmark{2,1,3,6}, 
Tatsuya \textsc{Nakaoka}\altaffilmark{1}, 
Katsutoshi \textsc{Takaki}\altaffilmark{1}, 
Daiki \textsc{Fukushima}\altaffilmark{7}, 
Naoto \textsc{Kojiguchi}\altaffilmark{7}, 
Kazunari \textsc{Masumoto}\altaffilmark{7}, 
Katsura \textsc{Matsumoto}\altaffilmark{7}, 
Hiroshi \textsc{Akitaya}\altaffilmark{8,2,1,3}, 
Ryosuke \textsc{Itoh}\altaffilmark{9,1}, 
Yuka \textsc{Kanda}\altaffilmark{1}, 
Yuki \textsc{Moritani}\altaffilmark{5,2,1}, 
Koji \textsc{Takata}\altaffilmark{1}, 
Makoto \textsc{Uemura}\altaffilmark{2,1,3}, 
Takahiro \textsc{Ui}\altaffilmark{1}, 
Michitoshi \textsc{Yoshida}\altaffilmark{10,2,1,3}, 
Takashi \textsc{Hattori}\altaffilmark{10}, 
%%%Kentaro \textsc{Aoki}\altaffilmark{8}, 
Chien-Hsiu \textsc{Lee}\altaffilmark{10}, 
Nozomu \textsc{Tominaga}\altaffilmark{6,5}, 
and Ken'ichi \textsc{Nomoto}\altaffilmark{5}.}
\altaffiltext{1}{Department of Physical Science, Hirosima University, Kagamiyama, Higashi-Hiroshima, Hiroshima 739-8526, Japan; kawabata@astro.hiroshima-u.ac.jp}
\altaffiltext{2}{Hirosima Astrophysical Science Center, Hiroshima University, Kagamiyama, Higashi-Hiroshima, Hiroshima 739-8526, Japan}
\altaffiltext{3}{Core Research for Energetic Universe (CORE-U), Hiroshima University, Kagamiyama, Higashi-Hiroshima, Hiroshima 739-8526, Japan}
\altaffiltext{4}{Department of Astronomy, Graduate School of Science, Kyoto University, Kitashirakawa-Oikawacho, Sakyo-ku, Kyoto 606-8502, Japan}
\altaffiltext{5}{Kavli Institute for the Physics and Mathematics of the Universe (WPI), The University of Tokyo, 5-1-5 Kashiwanoha, Kashiwa, Chiba 277-8583, Japan}
\altaffiltext{6}{Department of Physics, Faculty of Science and Engineering, Konan University, Okamoto, Kobe, Hyogo 658-8501, Japan}
\altaffiltext{7}{Astronomical Institute, Osaka Kyoiku University, Asahigaoka, Kashiwara, Osaka 582-8582, Japan}
\altaffiltext{8}{Graduate School of Science and Engineering, Saitama University, 255 Shimo-Okubo, Sakura-ku, Saitama, 338-8570, Japan}
\altaffiltext{9}{School of Science, Tokyo Institute of Technology, 2-12-1 Ohokayama, Meguro-ku, Tokyo 152-8551, Japan}
\altaffiltext{10}{Subaru Telescope, National Astronomical Observatory of Japan, 650 North A'ohoku Place, Hilo, HI 96720 USA}
%\altaffiltext{11}{Hamamatsu Professor}
\maketitle

\begin{abstract}
We present optical and near-infrared observations of 
the nearby Type Iax supernova (SN) 2014dt from $14$ to 
$410$ days after the maximum light.
The velocities of the iron absorption lines in the early 
phase indicated that SN 2014dt showed slower expansion than 
the well-observed Type Iax SNe 2002cx, 2005hk and 2012Z.
In the late phase, the evolution of the light curve
and that of the spectra were considerably slower.
The spectral energy distribution kept roughly the same 
shape after $\sim 100$ days, and the bolometric light curve 
flattened during the same period.
These observations suggest the existence of an optically 
thick component that almost fully trapped the $\gamma$-ray 
energy from $^{56}$Co decay.
These findings are consistent with the predictions of 
the weak deflagration model, leaving a bound white dwarf 
remnant after the explosion.
\end{abstract}

\section{Introduction}

For normal Type Ia supernovae (SNe Ia), there is a 
well-established correlation between the peak luminosity 
and the decline rate of the light curve (LC; e.g., Phillips 
1993), which has been used to measure distances 
to remote galaxies. 
However, some SNe Ia deviate from this correlation, and 
their peak luminosities are significantly fainter ($\geq$ 1 
mag) than those of normal SNe Ia with similar decline rates.
These outliers have been called SN 2002cx-like SNe 
(Li et al. 2003) or SNe Iax (Foley et al. 2013).
SNe Iax commonly show lower luminosities, lower expansion 
velocities and hotter photospheres around their maximum 
light than normal SNe Ia (Foley et al. 2013 and references 
therein), as well as scatter in maximum magnitudes 
and decline rates.
For example, the maximum magnitude of the faintest SN Iax 
2008ha is only $-13.74 \pm 0.15$ mag in the $B$-band 
(Foley et al. 2009), which is $\sim$ 4 mag fainter than 
that of the prototypical SN Iax SN 2002cx.
The line velocities of SNe Iax around the maximum light 
are $2000$--$8000$ km s$^{-1}$, whereas those of normal 
SNe Ia are $\gtrsim 10000$ km s$^{-1}$.
The estimated $^{56}$Ni mass and explosion energy of 
SNe Iax are only $0.003$--$0.3$ M$_{\odot}$ and 
$10^{49}$--$10^{51}$ erg, respectively 
(e.g., Foley et al. 2009, Foley et al. 2013, Stritzinger 
et al. 2015).

From the compilation of early phase observations of SNe Iax, 
weak correlations between the maximum magnitude and the 
decline rate, line velocity and rising time have been 
suggested (e.g., Narayan et al. 2011; Foley et al. 2013; 
Magee et al. 2016).
However, rapidly increasing observational 
data suggest that this is probably an oversimplification 
and that SNe Iax show large diversity as well.
For example, SN Iax 2014ck showed spectra similar to SN 
2008ha, whereas the LCs were similar to SNe 2002cx and 
2005hk (Tomasella et al. 2016).
PS1-12bwh showed a spectral evolution similar to that of 
SN 2005hk in the post-maximum phase, whereas the 
pre-maximum evolution showed significant difference 
(Magee et al. 2017).
SN 2012Z showed close similarity to SN 2005hk 
in its early phase spectra.
In the late phase, although the emission lines of SN 2012Z 
were significantly broader than those of SN 2005hk, the overall 
spectral features of SN 2012Z were similar to those of SN 2005hk. 
Regardless of the similarity seen in the spectra, the decline 
rates of the LCs in the late phase show the diversity (Stritzinger 
et al. 2015, Yamanaka et al. 2015).
 
These peculiarities of SNe Iax are difficult to explain 
within the framework of canonical SNe Ia.
Many researchers have discussed plausible models that can 
reproduce the observed properties.
One of the models is the weak deflagration of a 
Chandrasekhar-mass carbon-oxygen white dwarf (e.g., Fink 
et al. 2014), in which a considerable part of the white 
dwarf could not gain sufficient kinetic energy to exceed 
the binding energy, and a bound white dwarf remnant of 
$\sim$ 1 M$_{\odot}$ may be left after the explosion.
However, Valenti et al. (2009) suggested that a fraction 
of SNe Iax could be core-collapse (CC) SNe based on 
of some characteristics in the optical spectra.
Moriya et al. (2010) suggested a `fallback CC SN model' 
for SNe Iax, where a considerable fraction of the ejecta 
would fall onto the central remnant, probably a black hole, 
in a CC SN explosion of a massive star 
($\gtrsim$ 25 M$_{\odot}$).

To date, the number of SNe Iax for which the evolution of 
the late phases has been well-observed remains small.
Late phase observations are expected to carry important 
information on the progenitor and explosion through a 
direct view to the inner part of the ejecta (e.g., Maeda 
et al. 2010).
The spectra of some SNe Iax in the late phase show the 
permitted lines with the P-Cygni profile, unlike those of 
SNe Ia.
From the analysis of the late phase spectra, Foley et al. 
(2016) suggested that the ejecta of SNe Iax have two 
distinct components, one of which creates the photosphere 
even in the late phase.
Thus, the data for SNe Iax in the late phases can provide 
meaningful constraints on an explosion model.
In addition, the low expansion velocity in SNe Iax makes it 
easier to identify different subclasses than in SNe Ia due 
to the smaller contamination by line blending (e.g., 
Stritzinger et al. 2014; but see Szalai et al. 2015 for 
practical difficulties in this kind of analysis).

The recent apparent identification of possible progenitor 
systems for two SNe Iax could be important for exploring 
the explosion model and links to normal SNe Ia.
For SN 2012Z, a blue hot star (possibly a companion helium 
star) has been detected in pre-explosion images 
(McCully et al. 2014).
For SN 2008ha, a luminous redder source has appeared in the 
post-explosion images, which could be a bound remnant 
or a survived companion star (Foley et al. 2014).
For other SN Iax (SN 2008ge; Foley et al. 2010 and 
SN 2014ck; Tomasella et al. 2016), only the upper limits 
for the brightness of the progenitors have been derived, 
excluding a massive progenitor scenario.

SN 2014dt was discovered by K. Itagaki in a nearby galaxy 
M61 on 2014 October 29.8 (UT) at 13.6 mag (Nakano et al. 
2014). 
The spectrum obtained on October 31, 2014, by the Asiago 
Supernova classification program is consistent with a SN 
Iax at $\sim$ 1 week after the maximum light 
(Ochner et al. 2014).
SN 2014dt is one of the nearest SN Iax ever discovered 
(see below).
Foley et al. (2015) gave an upper limit for the luminosity 
of the progenitor system using pre-explosion images taken 
with the Hubble Space Telescope (HST), which is consistent 
with a system with a low-mass evolved or main-sequence 
companion.
However, the upper limit could not reject the system with 
a sub-class of Wolf-Rayet star.
The distance of M61 obtained by several works are summarized 
in the NASA/IPAC Extragalactic Database (NED).
The Tully-Fisher relation (Schoeniger \& Sofue 1997) shows 
a large scatter ($\mu = 30.02 \pm 0.92$ mag using the CO 
Tully-Fisher relation, and $\mu = 30.21 \pm 0.70$ 
mag using the H{\sc i} Tully-Fisher relation), thus we decided 
not to adopt these values.
Bose \& Kumar et al. (2014) reported the distance obtained by 
applying the expanding photosphere method (EPM) to SN 2008in 
(Bose \& Kumar et al. 2014).
They gave two possible values for the distance with two 
different atmosphere models, $\mu = 30.45 \pm 0.10$ mag and 
$\mu = 30.81 \pm 0.20$ mag.
In this paper, we adopt the EPM distance of M61, 
$\mu = 30.81 \pm 0.20$ mag, which was calibrated with the 
atmosphere model updated by Dessart \& Hillier (2005).
Some parameters (e.g., the luminosity, the $^{56}$Ni mass) 
are affected by the difference of the adopted distance.
In this paper we adopt the EPM distance, while the parameter 
values derived with the Tully-Fisher distances will be 
mentioned as well to show associated uncertainty.
For the total extinction towards SN 2014dt, we simply 
adopted $E(B-V)_{\rm total} = 0.02$ mag, following 
Foley et al. (2015).

In this paper, we report our extended optical and 
near-infrared (NIR) observations of SN 2014dt.
We describe the observation and data reduction in 
\S 2.
In \S 3, we present the results of our observations.
In \S 4, we investigate the characteristic 
features of SN 2014dt and discuss the implications for the 
progenitor and explosion mechanism. 
In the analyses, we emphasize the characteristic behaviors 
found in the late phase. 
The slow evolution seen in the LCs and spectral energy 
distribution (SED) suggests the existence of 
high-density material buried deep in the ejecta. 
A similar suggestion was made by Foley et al. (2016), based 
on late phase spectra, but our conclusions were reached by 
independent observational materials.
A summary of this work is provided in \S 5.

\section{Observations and Data Reduction}

We performed optical $BVRI$-band photometry of SN 2014dt 
using Hiroshima One-shot Wide-field Polarimeter (HOWPol; 
Kawabata et al. 2008) installed at the 1.5 m Kanata telescope
at Higashi-Hiroshima Observatory and with an Andor DW936N-BV
CCD camera installed at the 51 cm telescope at Osaka Kyoiku 
University (OKU).
We reduced the data in a standard manner for CCD photometry.
For the images obtained by HOWPol, we performed subtraction 
of the host galaxy template image before aperture 
photometry to minimize contamination by the host galaxy 
background (Figure \ref{14dt_sub}).
The template images for the $BVRI$-bands were obtained with 
the same telescope and instrument on 2018 February 23 (i.e., 
$> 1200$ days after the discovery; Figure 2).
We removed the cosmic rays using {\it L. A. Cosmic} pipeline 
(van Dokkum 2001, van Dokkum et al. 2012).
After that, we remapped our data to WCS using the {\it SExtractor} 
(Bertin \& Arnouts 1996), and transformed the template images 
into our images using $wcsremap$\footnotemark[1]\footnotetext[1]{
http://www.astro.washington.edu/users/becker/v2.0/wcsremap.html}.
The host galaxy subtraction was then performed using $hotpants$\footnotemark[2]
\footnotetext[2]{http://www.astro.washington.edu/users/becker/v2.0/hotpants.html}.
For the subsequent aperture photometry, 
we used the DAOPHOT package in {\it IRAF}\footnotemark[3].
\footnotetext[3]{\ {\it IRAF} is 
distributed by the National Optical Astronomy Observatory, 
which is operated by the Association of Universities for 
Research in Astronomy (AURA) under a cooperative agreement 
with the National Science Foundation.}
We estimated the error for the photometry from the error 
associated with photometry, the variation in the magnitude 
obtained from each comparison stars and the difference images.
The error for the photometry included that caused by the 
host galaxy subtraction.
The estimated error due to the host galaxy subtraction 
is dominant.

For the images obtained by Andor CCD camera on the 51cm 
telescope, we skipped the template subtraction 
because the images were significantly affected by fringe 
patterns and the template subtraction made the photometric 
results worse.
We, however, note that the host galaxy contamination is 
negligible for those images, as those were obtained in the 
early phases when the SN itself was much brighter than the 
host background (as was confirmed by the subtracted images 
for the HOWPol data).
We adopted Point-Spread-Function (PSF) fitting photometry 
for these data.

\begin{figure}
  \begin{center}
   \includegraphics[width=80mm,clip,bb=0 0 680 900]{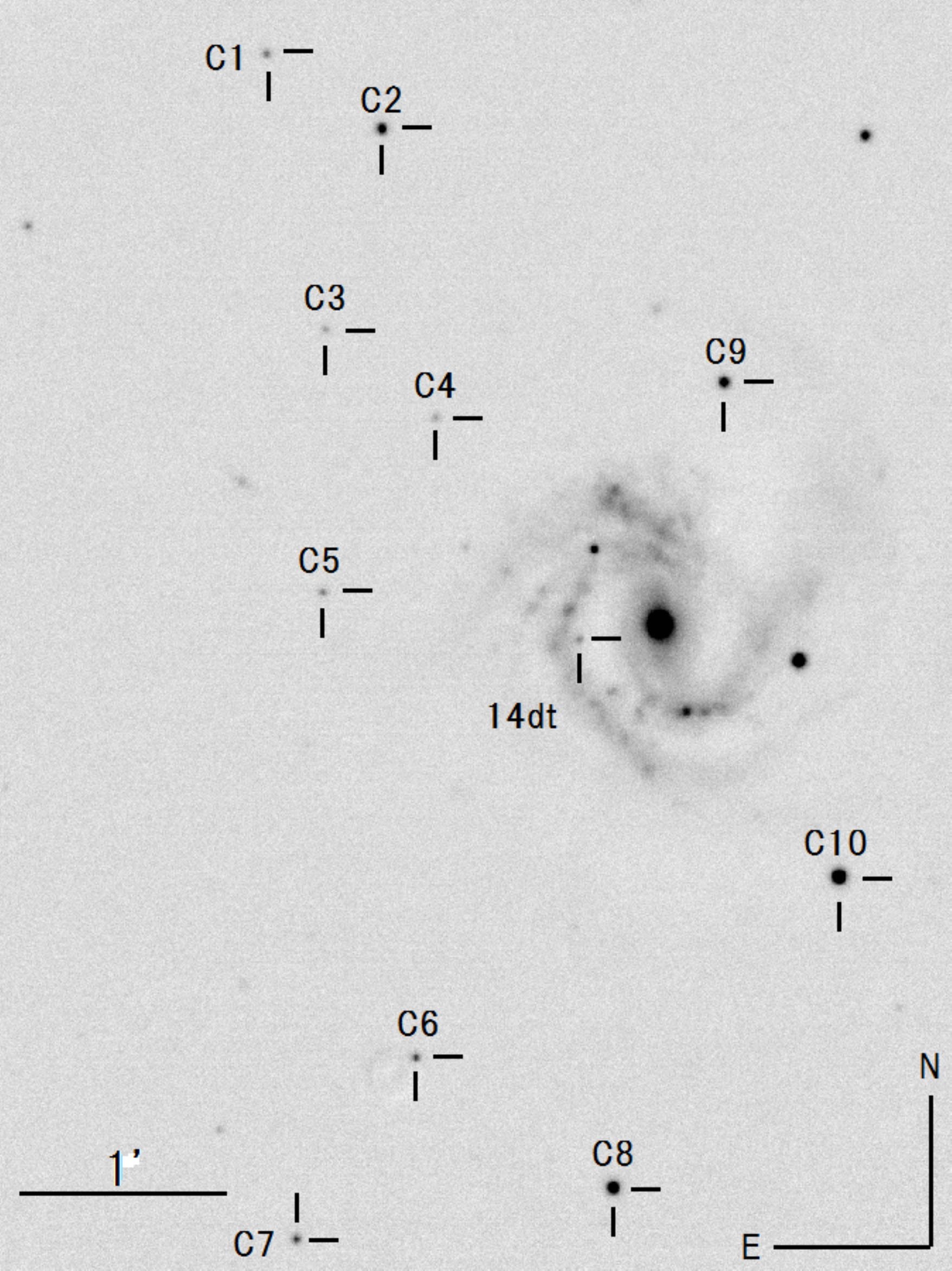}
  \end{center}
  \caption{The $R$-band image of SN 2014dt and comparison stars 
taken with the Kanata telescope/HOWPol on MJD 57171.51 
(2015 March 29).
}\label{14dt_fov}
\end{figure}

\begin{figure}
  \begin{center}
   \includegraphics[width=80mm,clip,bb=0 0 440 400]{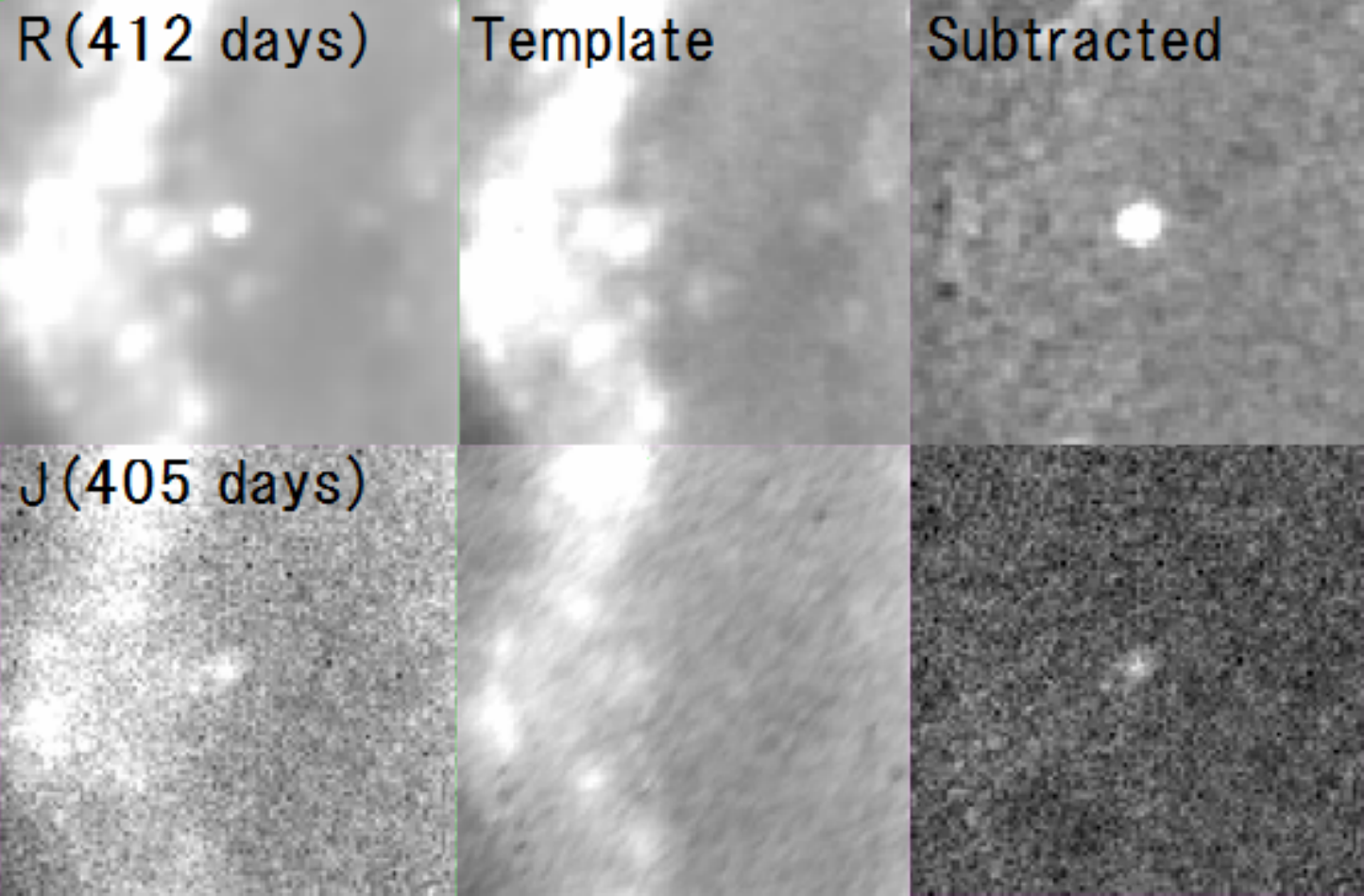}
  \end{center}
  \caption{Comparison of the observed (left panels), the 
template (center), and the subtracted (right) images around 
SN 2014dt. 
Upper and lower panels are images in the $R$-band at 412 
days and in the $J$-band at 405 days, respectively. 
The field of view is $30''\times 30''$.
The template image in the $R$-band is synthesized from SDSS images, 
and that in the $J$-band was obtained with HONIR (see text).
}\label{14dt_sub}
\end{figure}

For the magnitude calibration, we adopted 
relative photometry using comparison stars taken in the 
same frame (Figure \ref{14dt_fov}).
The magnitudes of the comparison stars were calibrated with
the photometric standard stars in the Landolt field (SA104; 
Landolt 1992), observed on photometric nights using HOWPol.
We show the magnitudes of the comparison stars in Table 1.
The magnitudes are consistent with those reported by Roy 
et al. (2011). 
We note that the $B$-band magnitudes of the comparison stars, 
as well as those given by Roy et al. (2011), are $\sim$0.3 mag 
fainer those shows given by Singh et al. (2018), perhaps due 
to a systematic difference between the systems.
For photometric correction, we performed first-order color 
term correction.
We skipped S-correction, which is negligible for the 
purposes of this study (Stritzinger et al. 2002).
We list the journal of the optical photometry in Table 2.

We performed NIR $JHK_{s}$-band photometry with Hiroshima 
Optical and Near-InfraRed camera (HONIR; Akitaya et al. 
2014) installed at the 1.5 m Kanata telescope.
In the observations, we adopted five-position dithering 
method.
We adopted the host galaxy subtraction and the aperture 
photometry method in the same way as used for reduction of 
the optical data, and calibrated the magnitude 
of the SN using the comparison stars in the 2MASS catalog 
(Skrutskie et al. 2006).
We obtained the galaxy template images with the same 
telescope and instrument on 2017 April 23 (i.e., at $> 900$ 
days after the discovery; Figure \ref{14dt_sub}).
For the photometric error, we included the error caused by 
the host galaxy in the same way as for the HOWPol data.
The journal of NIR photometry is shown in Table 3.

\begin{longtable}{lccccccc}
\caption{Magnitudes of comparison stars of SN 2014dt}
\hline
ID & $B$ (mag) & $V$ (mag) & $R$ (mag) & $I$ (mag) & $J$ (mag)\footnotemark[4] & $H$ (mag)\footnotemark[4] & $K_{s}$ (mag)\footnotemark[4] \\
\hline
\endhead
\hline
\endfoot
\hline
\multicolumn{4}{l}{\hbox to 0pt{\parbox{180mm}{\footnotesize
\footnotemark[4] The magnitudes in the NIR bands are from 
the 2MASS catalog (Skrutskie et al. 2006).
}\hss}}
\endlastfoot
   C1 & 17.536 $\pm$ 0.026 & 16.842 $\pm$ 0.012 & 16.338 $\pm$ 0.024 & 15.992 $\pm$ 0.012 &  $-$  &  $-$  &  $-$  \\ \hline
   C2 & 16.340 $\pm$ 0.023 & 15.505 $\pm$ 0.010 & 14.970 $\pm$ 0.034 & 14.533 $\pm$ 0.016 &  $-$  &  $-$  &  $-$  \\ \hline
   C3 & 18.451 $\pm$ 0.033 & 17.682 $\pm$ 0.015 & 17.170 $\pm$ 0.025 & 16.812 $\pm$ 0.014 &  $-$  &  $-$  &  $-$  \\ \hline
   C4 & 19.585 $\pm$ 0.066 & 18.053 $\pm$ 0.019 & 16.971 $\pm$ 0.025 & 16.061 $\pm$ 0.012 & 15.261 $\pm$ 0.055 & 14.571 $\pm$ 0.070 & 14.377 $\pm$ 0.099 \\ \hline
   C5 & 18.785 $\pm$ 0.039 & 17.552 $\pm$ 0.014 & 16.575 $\pm$ 0.024 & 15.872 $\pm$ 0.012 & 15.135 $\pm$ 0.059 & 14.598 $\pm$ 0.051 & 14.343 $\pm$ 0.087 \\ \hline
   C6 & 18.280 $\pm$ 0.032 & 16.956 $\pm$ 0.012 & 15.977 $\pm$ 0.024 & 15.252 $\pm$ 0.011 & 14.619 $\pm$ 0.042 & 13.956 $\pm$ 0.038 & 13.831 $\pm$ 0.061 \\ \hline
   C7 & 18.256 $\pm$ 0.032 & 16.895 $\pm$ 0.012 & 15.878 $\pm$ 0.024 & 15.144 $\pm$ 0.011 &  $-$  &  $-$  &  $-$  \\ \hline
   C8 & 15.542 $\pm$ 0.023 & 14.731 $\pm$ 0.010 & 14.164 $\pm$ 0.024 & 13.748 $\pm$ 0.011 & 13.525 $\pm$ 0.026 & 13.115 $\pm$ 0.028 & 13.053 $\pm$ 0.035 \\ \hline
   C9 & 16.844 $\pm$ 0.024 & 15.609 $\pm$ 0.010 & 14.719 $\pm$ 0.024 & 14.048 $\pm$ 0.011 & 13.326 $\pm$ 0.023 & 12.724 $\pm$ 0.026 & 12.602 $\pm$ 0.033 \\ \hline
   C10 &   $-$  &  $-$  &  $-$  &  $-$  & 12.734 $\pm$ 0.026 & 12.388 $\pm$ 0.030 & 12.352 $\pm$ 0.031 \\ \hline
\end{longtable}

We obtained optical spectra of SN 2014dt using HOWPol 
at 13 epochs.
The wavelength coverage was $4500$--$9200$ \AA \ and the 
wavelength resolution was $R = \lambda/\Delta\lambda \simeq 400$ 
at 6000 \AA.
For wavelength calibration, we used sky emission lines.
To remove cosmic ray events, we used the 
{\it L. A. Cosmic} pipeline. 
The flux of SN 2014dt was calibrated using data of 
spectrophotometric standard stars taken on the same night.
The journal of spectroscopy is listed in Table 4.

In addition, we obtained optical photometric and spectroscopic
data in the late phases with the Faint Object Camera and 
Spectrograph (FOCAS; Kashikawa et al. 2002) installed at 
the 8.2 m Subaru telescope, NAOJ.
The wavelength resolution was $R = \lambda/\Delta\lambda \simeq 650$ 
at 6000 \AA.
The data reduction was performed in the same way as that 
with HOWPol, except for the host galaxy subtraction for 
the imaging photometry and the wavelength calibration for 
the spectroscopy.
The template images for the $V$ and $R$-bands were created 
from the SDSS $u'g'r'i'z'$ images using the relations for 
the flux transformation (Smith et al. 2002).
For the wavelength calibration, we used arc lamp (Th-Ar) 
data and skylines.
The observation log with Subaru/FOCAS is shown in Tables 
2 and 4.

\begin{longtable}{lccccccc}
\caption{Log of optical photometry of SN 2014dt}
\hline
Date & MJD & Phase\footnotemark[5] & $B$ (mag) & $V$ (mag) & $R$ (mag) & $I$ (mag) & Site\footnotemark[6]\\
\hline
\endhead
\hline
\endfoot
\hline
\multicolumn{7}{l}{\hbox to 0pt{\parbox{180mm}{\footnotesize
\footnotemark[5] The uncertainty of phase is inevitably larger, $\pm 4$ days. See \S3.3.
   \par\noindent
\footnotemark[6] See \S2.
}\hss}}
\endlastfoot
    2014-11-03 & 56964.85 & 14.4 &  $-$  &  $-$  & 13.43 $\pm$ 0.03 & 13.31 $\pm$ 0.02 & OKU \\
    2014-11-04 & 56965.84 & 15.4 &  $-$  &  $-$  & 13.51 $\pm$ 0.04 & 13.37 $\pm$ 0.01 & OKU \\
    2014-11-05 & 56966.83 & 16.4 &       $-$          &        $-$         & 13.654 $\pm$ 0.051 & 13.313 $\pm$ 0.050 & Kanata \\
    2014-11-06 & 56967.86 & 17.5 &       $-$          & 14.349 $\pm$ 0.054 & 13.743 $\pm$ 0.055 & 13.312 $\pm$ 0.051 & Kanata \\
    2014-11-10 & 56971.84 & 22.4 & 16.113 $\pm$ 0.050 & 14.522 $\pm$ 0.050 & 13.965 $\pm$ 0.051 & 13.580 $\pm$ 0.051 & Kanata \\
    2014-11-10 & 56971.85 & 22.4 &  $-$  & 14.65 $\pm$ 0.07 & 13.97 $\pm$ 0.04 & 13.63 $\pm$ 0.04 & OKU \\
    2014-11-13 & 56974.84 & 24.4 &       $-$          &        $-$         & 14.165 $\pm$ 0.050 & 13.894 $\pm$ 0.120 & Kanata \\
    2014-11-16 & 56977.83 & 27.4 &  $-$  & 14.95 $\pm$ 0.03 & 14.23 $\pm$ 0.05 & 13.81 $\pm$ 0.03 & OKU \\
    2014-11-17 & 56978.86 & 28.5 &  $-$  &  $-$  & 14.34 $\pm$ 0.03 & 14.03 $\pm$ 0.02 & OKU \\
    2014-11-18 & 56979.85 & 29.4 & 16.54 $\pm$ 0.04 & 14.98 $\pm$ 0.02 & 14.35 $\pm$ 0.03 & 14.02 $\pm$ 0.02 & OKU \\
    2014-11-18 & 56979.85 & 29.4 & 16.594 $\pm$ 0.054 & 15.125 $\pm$ 0.051 & 14.380 $\pm$ 0.050 & 13.964 $\pm$ 0.050 & Kanata \\
    2014-11-19 & 56980.84 & 30.4 &  $-$  & 15.02 $\pm$ 0.08 &  $-$  &  $-$  & OKU \\
    2014-11-20 & 56981.84 & 31.4 & 16.68 $\pm$ 0.06 & 15.02 $\pm$ 0.03 & 14.42 $\pm$ 0.03 & 14.12 $\pm$ 0.02 & OKU \\
    2014-11-22 & 56983.84 & 33.4 & 16.846 $\pm$ 0.050 &        $-$         &        $-$         &        $-$         & Kanata \\
    2014-11-22 & 56983.85 & 33.4 & 16.68 $\pm$ 0.07 & 15.11 $\pm$ 0.01 & 14.54 $\pm$ 0.03 & 14.11 $\pm$ 0.10 & OKU \\
    2014-11-23 & 56984.83 & 34.4 & 16.82 $\pm$ 0.08 & 15.14 $\pm$ 0.03 & 14.54 $\pm$ 0.03 & 14.23 $\pm$ 0.02 & OKU \\
    2014-11-27 & 56988.84 & 38.4 & 16.77 $\pm$ 0.06 & 15.25 $\pm$ 0.03 & 14.70 $\pm$ 0.03 & 14.37 $\pm$ 0.02 & OKU \\
    2014-11-29 & 56990.85 & 40.4 & 16.73 $\pm$ 0.06 & 15.33 $\pm$ 0.02 & 14.78 $\pm$ 0.03 & 14.48 $\pm$ 0.05 & OKU \\
    2014-11-29 & 56990.87 & 40.5 & 16.848 $\pm$ 0.050 &        $-$         &        $-$         &        $-$         & Kanata \\
    2014-12-05 & 56996.86 & 46.5 & 16.80 $\pm$ 0.03 & 15.44 $\pm$ 0.02 & 14.94 $\pm$ 0.05 &  $-$  & OKU \\
    2014-12-06 & 56997.80 & 47.4 & 17.10 $\pm$ 0.05 & 15.48 $\pm$ 0.03 & 15.03 $\pm$ 0.03 & 14.72 $\pm$ 0.02 & OKU \\
    2014-12-07 & 56998.86 & 48.5 & 16.87 $\pm$ 0.06 & 15.49 $\pm$ 0.01 & 15.03 $\pm$ 0.03 & 14.72 $\pm$ 0.02 & OKU \\
    2014-12-07 & 56998.86 & 48.5 & 17.020 $\pm$ 0.054 &        $-$         &        $-$         &        $-$         & Kanata \\
    2014-12-08 & 56999.84 & 49.4 & 16.90 $\pm$ 0.04 & 15.49 $\pm$ 0.03 &  $-$  &  $-$  & OKU \\
    2014-12-11 & 57002.80 & 52.4 & 17.08 $\pm$ 0.08 & 15.58 $\pm$ 0.06 & 15.13 $\pm$ 0.03 & 14.81 $\pm$ 0.01 & OKU \\
    2014-12-13 & 57004.81 & 54.4 & 16.98 $\pm$ 0.04 & 15.60 $\pm$ 0.02 & 15.17 $\pm$ 0.03 & 15.05 $\pm$ 0.31 & OKU \\
    2014-12-14 & 57005.86 & 55.5 & 17.09 $\pm$ 0.04 & 15.63 $\pm$ 0.02 & 15.19 $\pm$ 0.03 & 14.89 $\pm$ 0.02 & OKU \\
    2014-12-20 & 57011.86 & 61.5 & 17.068 $\pm$ 0.050 & 15.931 $\pm$ 0.052 & 15.425 $\pm$ 0.051 & 14.880 $\pm$ 0.051 & Kanata \\
    2014-12-21 & 57012.80 & 62.4 & 17.00 $\pm$ 0.08 & 15.81 $\pm$ 0.05 & 15.38 $\pm$ 0.04 & 15.06 $\pm$ 0.01 & OKU \\
    2014-12-22 & 57013.85 & 63.4 & 17.22 $\pm$ 0.21 & 15.67 $\pm$ 0.22 & 15.40 $\pm$ 0.03 & 15.01 $\pm$ 0.10 & OKU \\
    2014-12-23 & 57014.83 & 64.4 & 17.01 $\pm$ 0.03 & 15.79 $\pm$ 0.02 & 15.41 $\pm$ 0.04 & 15.10 $\pm$ 0.02 & OKU \\
    2014-12-24 & 57015.87 & 65.5 & 17.15 $\pm$ 0.11 & 15.86 $\pm$ 0.05 & 15.48 $\pm$ 0.06 &  $-$  & OKU \\
    2014-12-26 & 57017.78 & 67.4 & 16.98 $\pm$ 0.03 & 15.87 $\pm$ 0.02 & 15.47 $\pm$ 0.04 & 15.05 $\pm$ 0.08 & OKU \\
    2014-12-26 & 57017.87 & 68.4 & 17.099 $\pm$ 0.050 & 16.078 $\pm$ 0.051 & 15.541 $\pm$ 0.051 & 14.984 $\pm$ 0.050 & Kanata \\
    2014-12-27 & 57018.80 & 68.4 & 17.18 $\pm$ 0.03 & 15.84 $\pm$ 0.03 & 15.48 $\pm$ 0.03 & 15.20 $\pm$ 0.01 & OKU \\
    2014-12-29 & 57020.80 & 70.4 & 17.18 $\pm$ 0.06 & 15.93 $\pm$ 0.01 & 15.56 $\pm$ 0.03 & 15.26 $\pm$ 0.01 & OKU \\
    2014-12-30 & 57021.83 & 71.4 & 17.24 $\pm$ 0.04 & 15.96 $\pm$ 0.02 & 15.59 $\pm$ 0.03 & 15.29 $\pm$ 0.02 & OKU \\
    2015-01-02 & 57024.86 & 74.4 & 17.27 $\pm$ 0.15 &  $-$  &  $-$  &  $-$  & OKU \\
    2015-01-03 & 57025.80 & 75.4 & 17.11 $\pm$ 0.15 &  $-$  &  $-$  &  $-$  & OKU \\
    2015-01-04 & 57026.83 & 76.4 & 17.34 $\pm$ 0.04 & 16.02 $\pm$ 0.02 & 15.70 $\pm$ 0.03 &  $-$  & OKU \\
    2015-01-08 & 57030.86 & 80.5 & 17.36 $\pm$ 0.04 &  $-$  &  $-$  &  $-$  & OKU \\
    2015-01-10 & 57032.80 & 82.4 & 17.49 $\pm$ 0.07 & 16.03 $\pm$ 0.12 & 15.80 $\pm$ 0.03 & 15.41 $\pm$ 0.02 & OKU \\
    2015-01-10 & 57032.83 & 82.4 & 17.778 $\pm$ 0.054 & 16.334 $\pm$ 0.051 & 16.002 $\pm$ 0.051 & 15.348 $\pm$ 0.050 & Kanata \\
    2015-01-16 & 57035.79 & 85.4 & 17.25 $\pm$ 0.07 & 16.14 $\pm$ 0.03 & 15.86 $\pm$ 0.03 & 15.47 $\pm$ 0.02 & OKU \\
    2015-01-17 & 57038.76 & 88.4 & 17.41 $\pm$ 0.04 &  $-$  &  $-$  &  $-$  & OKU \\
    2015-01-20 & 57039.82 & 89.4 & 17.32 $\pm$ 0.04 & 16.32 $\pm$ 0.06 & 15.93 $\pm$ 0.04 & 15.58 $\pm$ 0.02 & OKU \\
    2015-01-23 & 57042.76 & 92.4 & 17.42 $\pm$ 0.05 & 16.25 $\pm$ 0.03 & 15.96 $\pm$ 0.11 & 15.34 $\pm$ 0.06 & OKU \\
    2015-01-24 & 57045.82 & 96.3 & 17.34 $\pm$ 0.06 & 16.31 $\pm$ 0.02 & 16.06 $\pm$ 0.06 &  $-$  & OKU \\
    2015-01-24 & 57046.67 & 96.3 & 17.506 $\pm$ 0.050 & 16.665 $\pm$ 0.052 &        $-$         & 15.484 $\pm$ 0.051 & Kanata \\
    2015-01-24 & 57046.69 & 96.3 & 17.51 $\pm$ 0.05 & 16.34 $\pm$ 0.02 & 16.08 $\pm$ 0.04 &  $-$  & OKU \\
    2015-01-28 & 57050.71 & 100.3 & 17.31 $\pm$ 0.16 & 16.35 $\pm$ 0.04 & 15.99 $\pm$ 0.09 &  $-$  & OKU \\
    2015-01-31 & 57053.77 & 103.4 & 17.61 $\pm$ 0.03 & 16.44 $\pm$ 0.02 & 16.20 $\pm$ 0.05 &  $-$  & OKU \\
    2015-02-01 & 57054.75 & 104.3 & 17.57 $\pm$ 0.04 &  $-$  &  $-$  &  $-$  & OKU \\
    2015-02-01 & 57054.76 & 104.4 & 17.560 $\pm$ 0.059 & 16.699 $\pm$ 0.056 &        $-$         &        $-$         & Kanata \\
    2015-02-06 & 57059.72 & 109.3 &  $-$  & 16.61 $\pm$ 0.04 & 16.31 $\pm$ 0.03 &  $-$  & OKU \\
    2015-02-10 & 57063.71 & 113.3 & 17.74 $\pm$ 0.04 & 16.61 $\pm$ 0.02 &  $-$  &  $-$  & OKU \\
    2015-02-11 & 57064.75 & 114.3 &  $-$  &  $-$  & 16.22 $\pm$ 0.07 &  $-$  & OKU \\
    2015-02-12 & 57065.75 & 115.3 & 17.68 $\pm$ 0.03 & 16.51 $\pm$ 0.12 & 16.40 $\pm$ 0.13 &  $-$  & OKU \\
    2015-02-13 & 57066.69 & 116.3 & 17.744 $\pm$ 0.050 & 16.793 $\pm$ 0.051 &        $-$         &        $-$         & Kanata \\
    2015-02-13 & 57066.81 & 116.4 & 17.77 $\pm$ 0.09 & 16.61 $\pm$ 0.03 & 16.29 $\pm$ 0.07 &  $-$  & OKU \\
    2015-02-14 & 57067.74 & 117.3 & 17.75 $\pm$ 0.03 & 16.65 $\pm$ 0.02 & 16.39 $\pm$ 0.04 &  $-$  & OKU \\
    2015-02-19 & 57072.67 & 122.3 & 17.65 $\pm$ 0.13 & 16.64 $\pm$ 0.06 & 16.40 $\pm$ 0.03 &  $-$  & OKU \\
    2015-02-20 & 57073.72 & 123.3 & 17.77 $\pm$ 0.05 & 16.68 $\pm$ 0.03 & 16.38 $\pm$ 0.05 &  $-$  & OKU \\
    2015-02-23 & 57076.70 & 126.3 & 17.90 $\pm$ 0.03 & 16.73 $\pm$ 0.03 & 16.50 $\pm$ 0.05 &  $-$  & OKU \\
    2015-02-24 & 57077.67 & 127.3 & 17.54 $\pm$ 0.05 & 16.50 $\pm$ 0.03 & 16.19 $\pm$ 0.05 &  $-$  & OKU \\
    2015-02-26 & 57079.83 & 129.4 & 18.120 $\pm$ 0.050 & 17.101 $\pm$ 0.054 & 16.615 $\pm$ 0.053 & 15.889 $\pm$ 0.053 & Kanata \\
    2015-03-01 & 57082.68 & 132.3 & 17.864 $\pm$ 0.050 & 17.169 $\pm$ 0.059 & 16.637 $\pm$ 0.058 & 15.845 $\pm$ 0.052 & Kanata \\
    2015-03-01 & 57082.74 & 132.3 & 17.56 $\pm$ 0.27 & 16.86 $\pm$ 0.15 & 16.48 $\pm$ 0.08 &  $-$  & OKU \\
    2015-03-02 & 57083.72 & 133.3 &  $-$  & 16.83 $\pm$ 0.02 & 16.47 $\pm$ 0.05 &  $-$  & OKU \\
    2015-03-04 & 57085.78 & 135.4 &  $-$  & 16.82 $\pm$ 0.08 &  $-$  &  $-$  & OKU \\
    2015-03-10 & 57091.65 & 141.2 & 17.76 $\pm$ 0.23 & 16.87 $\pm$ 0.04 & 16.61 $\pm$ 0.06 &  $-$  & OKU \\
    2015-03-11 & 57092.69 & 142.3 & 18.03 $\pm$ 0.03 & 16.98 $\pm$ 0.03 &  $-$  &  $-$  & OKU \\
    2015-03-12 & 57093.60 & 143.2 &  $-$  & 16.97 $\pm$ 0.02 & 16.69 $\pm$ 0.04 &  $-$  & OKU \\
    2015-03-14 & 57095.66 & 145.3 &  $-$  & 16.90 $\pm$ 0.19 &  $-$  &  $-$  & OKU \\
    2015-03-16 & 57097.59 & 147.2 &  $-$  & 16.98 $\pm$ 0.02 & 16.68 $\pm$ 0.05 & 16.17 $\pm$ 0.03 & OKU \\
    2015-03-17 & 57098.59 & 148.2 &  $-$  & 17.05 $\pm$ 0.05 & 16.67 $\pm$ 0.06 & 16.30 $\pm$ 0.27 & OKU \\
    2015-03-20 & 57101.62 & 151.2 &  $-$  & 16.97 $\pm$ 0.04 & 16.71 $\pm$ 0.03 & 16.19 $\pm$ 0.04 & OKU \\
    2015-03-21 & 57102.53 & 152.1 &       $-$          & 17.238 $\pm$ 0.056 & 16.757 $\pm$ 0.052 & 15.955 $\pm$ 0.052 & Kanata \\
    2015-03-21 & 57102.62 & 152.2 &  $-$  & 17.03 $\pm$ 0.07 & 16.75 $\pm$ 0.07 & 16.19 $\pm$ 0.11 & OKU \\
    2015-03-22 & 57103.69 & 153.3 &  $-$  & 17.05 $\pm$ 0.08 & 16.68 $\pm$ 0.05 & 16.19 $\pm$ 0.06 & OKU \\
    2015-03-23 & 57104.59 & 154.2 &  $-$  & 17.00 $\pm$ 0.08 & 16.63 $\pm$ 0.05 & 16.14 $\pm$ 0.02 & OKU \\
    2015-03-24 & 57105.58 & 155.2 &  $-$  & 17.07 $\pm$ 0.11 & 16.75 $\pm$ 0.05 & 16.21 $\pm$ 0.04 & OKU \\
    2015-03-25 & 57106.56 & 156.2 &  $-$  & 17.13 $\pm$ 0.05 & 16.75 $\pm$ 0.06 & 16.25 $\pm$ 0.03 & OKU \\
    2015-03-26 & 57107.53 & 157.1 &  $-$  & 17.12 $\pm$ 0.08 & 16.82 $\pm$ 0.04 & 16.33 $\pm$ 0.04 & OKU \\
    2015-03-27 & 57108.55 & 158.1 &  $-$  & 17.01 $\pm$ 0.13 & 16.73 $\pm$ 0.07 & 16.11 $\pm$ 0.14 & OKU \\
    2015-03-28 & 57109.57 & 159.2 & 18.28 $\pm$ 0.05 & 17.11 $\pm$ 0.08 & 16.81 $\pm$ 0.05 & 16.34 $\pm$ 0.05 & OKU \\
    2015-03-28 & 57109.60 & 159.2 &       $-$          &        $-$         & 16.529 $\pm$ 0.056 &        $-$         & Kanata \\
    2015-03-29 & 57110.61 & 160.2 & 18.297 $\pm$ 0.085 & 17.215 $\pm$ 0.062 & 16.787 $\pm$ 0.061 & 15.993 $\pm$ 0.056 & Kanata \\
    2015-03-30 & 57111.59 & 161.2 & 18.25 $\pm$ 0.12 & 17.07 $\pm$ 0.05 & 16.79 $\pm$ 0.03 & 16.30 $\pm$ 0.03 & OKU \\
    2015-04-11 & 57123.53 & 173.1 &  $-$  & 17.19 $\pm$ 0.09 & 16.87 $\pm$ 0.20 & 16.31 $\pm$ 0.11 & OKU \\
    2015-04-21 & 57133.53 & 183.1 & 18.780 $\pm$ 0.050 & 17.509 $\pm$ 0.078 & 17.127 $\pm$ 0.057 & 16.250 $\pm$ 0.052 & Kanata \\
    2015-05-05 & 57147.46 & 197.1 & 19.000 $\pm$ 0.050 & 17.629 $\pm$ 0.059 &        $-$         &        $-$         & Kanata \\
    2015-05-29 & 57171.51 & 221.1 & 19.571 $\pm$ 0.087 & 18.013 $\pm$ 0.052 & 17.269 $\pm$ 0.051 & 16.557 $\pm$ 0.050 & Kanata \\
    2015-06-06 & 57179.51 & 229.1 & 19.430 $\pm$ 0.050 & 17.887 $\pm$ 0.054 & 17.389 $\pm$ 0.054 & 16.444 $\pm$ 0.051 & Kanata \\
    2015-06-21 & 57194.26 & 243.9 &  $-$  & 17.959 $\pm$ 0.014 & 17.375 $\pm$ 0.014 &  $-$  & Subaru \\
    2015-06-28 & 57201.47 & 251.1 &        $-$         &        $-$         & 17.132 $\pm$ 0.059 &        $-$         & Kanata \\
    2015-06-29 & 57202.47 & 252.1 &        $-$         &        $-$         & 17.535 $\pm$ 0.063 &        $-$         & Kanata \\
    2015-08-01 & 57235.48 & 285.1 &        $-$         &        $-$         & 17.540 $\pm$ 0.066 & 16.726 $\pm$ 0.058 & Kanata \\
    2015-08-03 & 57237.47 & 287.1 &        $-$         &        $-$         & 17.623 $\pm$ 0.053 & 16.751 $\pm$ 0.054 & Kanata \\
    2015-08-05 & 57239.47 & 289.1 &        $-$         & 18.709 $\pm$ 0.091 & 17.596 $\pm$ 0.061 &        $-$         & Kanata \\
    2015-12-06 & 57362.59 & 412.2 &  $-$  & 19.355 $\pm$ 0.017 & 18.621 $\pm$ 0.011 &  $-$  & Subaru \\ \hline
\end{longtable}

\begin{longtable}{lccccc}
\caption{Log of NIR photometry SN 2014dt}
\hline
Date & MJD & Phase\footnotemark[7] & $J$ (mag) & $H$ (mag) & $K_{s}$ (mag) \\
\hline
\endhead
\hline
\endfoot
\hline
\multicolumn{4}{l}{\hbox to 0pt{\parbox{180mm}{\footnotesize
\footnotemark[7] The uncertainty of phase is inevitably larger, $\pm 4$ days. See \S3.3.
}\hss}}
\endlastfoot
    2014-11-22 & 56983.87 & 33.5  &        $-$         & 13.987 $\pm$ 0.300 & 14.415 $\pm$ 0.301 \\
    2014-11-29 & 56990.84 & 40.4  & 14.986 $\pm$ 0.301 & 14.090 $\pm$ 0.301 & 13.987 $\pm$ 0.301 \\
    2014-12-07 & 56998.87 & 48.5  & 15.224 $\pm$ 0.301 & 14.568 $\pm$ 0.301 & 14.870 $\pm$ 0.305 \\
    2015-01-24 & 57046.65 & 96.2  & 15.999 $\pm$ 0.306 &        $-$         &        $-$         \\
    2015-01-28 & 57050.80 & 100.4 & 15.925 $\pm$ 0.318 &        $-$         &        $-$         \\
    2015-02-01 & 57054.70 & 104.3 & 16.249 $\pm$ 0.305 & 15.635 $\pm$ 0.309 &        $-$         \\
    2015-02-08 & 57061.72 & 111.3 & 15.939 $\pm$ 0.300 &        $-$         &        $-$         \\
    2015-02-13 & 57066.66 & 116.3 & 16.294 $\pm$ 0.303 & 15.674 $\pm$ 0.302 &        $-$         \\
    2015-03-02 & 57083.63 & 132.2 & 16.309 $\pm$ 0.303 & 15.816 $\pm$ 0.303 &      $>$ 15.9      \\
    2015-03-10 & 57091.64 & 141.2 & 16.855 $\pm$ 0.336 & 15.939 $\pm$ 0.310 &      $>$ 14.7      \\
    2015-03-22 & 57103.58 & 153.2 & 16.271 $\pm$ 0.303 & 16.088 $\pm$ 0.310 &      $>$ 15.6      \\
    2015-03-25 & 57106.69 & 156.3 & 16.734 $\pm$ 0.308 &        $-$         &        $-$         \\
    2015-04-17 & 57129.64 & 179.2 &        $-$         &        $-$         &      $>$ 15.6      \\
    2015-04-25 & 57137.55 & 187.1 & 16.676 $\pm$ 0.309 & 16.427 $\pm$ 0.311 &        $-$         \\
    2015-05-02 & 57144.45 & 194.0 & 16.359 $\pm$ 0.300 &        $-$         &        $-$         \\
    2015-05-05 & 57147.49 & 197.1 & 16.614 $\pm$ 0.315 & 16.183 $\pm$ 0.321 &        $-$         \\
    2015-05-20 & 57162.61 & 212.2 & 16.727 $\pm$ 0.307 & 16.090 $\pm$ 0.316 &        $-$         \\
    2015-06-03 & 57176.47 & 226.1 & 16.993 $\pm$ 0.340 & 16.846 $\pm$ 0.326 &        $-$         \\
    2015-11-30 & 57355.82 & 405.4 & 16.991 $\pm$ 0.316 &        $-$         &        $-$         \\
    2015-12-20 & 57375.84 & 425.4 & 17.481 $\pm$ 0.348 & 17.025 $\pm$ 0.323 &      $>$ 17.5      \\ \hline
%    2015-12-20 & 57375.84 & 425.0 & 17.481 $pm$ 0.348 & 17.025 $pm$ 0.323 & 18.677 $pm$ 0.447 \\ \hline
\end{longtable}

\begin{longtable}{lcccccc}
\caption{Log of spectroscopic observations of SN 2014dt}
\hline
Date & MJD & Phase\footnotemark[8] & Telescope (Instrument) & Wavelength Range (\AA) & Resolution\footnotemark[9] & S/N\footnotemark[9] \\
\hline
\endhead
\hline
\endfoot
\hline
\multicolumn{4}{l}{\hbox to 0pt{\parbox{180mm}{\footnotesize
\footnotemark[8] The uncertainty of phase is inevitably larger, $\pm 4$ days. See \S3.3.
   \par\noindent
\footnotemark[9] At 6000\AA.
}\hss}}
\endlastfoot
   2014-11-10 & 56971.84 & 21.4 & Kanata (HOWPol)  & $4000$--$9000$  & 400 & 22 \\
   2014-11-18 & 56979.86 & 29.5 & Kanata (HOWPol)  & $4000$--$9000$  & 400 & 3  \\
   2014-11-29 & 56990.88 & 40.5 & Kanata (HOWPol)  & $4000$--$9000$  & 400 & 7  \\
   2014-12-23 & 57014.86 & 64.5 & Kanata (HOWPol)  & $4000$--$9000$  & 400 & 18 \\
   2015-01-10 & 57032.84 & 82.4 & Kanata (HOWPol)  & $4000$--$9000$  & 400 & 5  \\
   2015-01-24 & 57046.69 & 96.3 & Kanata (HOWPol)  & $4000$--$9000$  & 400 & 8  \\
   2015-02-01 & 57054.77 & 104.4 & Kanata (HOWPol) & $4000$--$9000$  & 400 & 9  \\
   2015-02-13 & 57066.70 & 116.3 & Kanata (HOWPol) & $4000$--$9000$  & 400 & 3  \\
   2015-02-21 & 57075.55 & 125.1 & Subaru (FOCAS)  & $3800$--$10000$ & 650 & 32 \\
   2015-03-02 & 57083.67 & 133.3 & Kanata (HOWPol) & $4000$--$9000$  & 400 & 3  \\
   2015-03-21 & 57102.55 & 152.1 & Kanata (HOWPol) & $4000$--$9000$  & 400 & 5  \\
   2015-04-15 & 57127.53 & 177.1 & Kanata (HOWPol) & $4000$--$9000$  & 400 & 2  \\
   2015-04-26 & 57138.53 & 188.1 & Kanata (HOWPol) & $4000$--$9000$  & 400 & 4  \\
   2015-05-29 & 57171.52 & 221.1 & Kanata (HOWPol) & $4000$--$9000$  & 400 & 1  \\
   2015-06-20 & 57194.26 & 243.6 & Subaru (FOCAS)  & $3800$--$10000$ & 650 & 32 \\
   2015-12-05 & 57362.59 & 412.2 & Subaru (FOCAS)  & $3800$--$10000$ & 650 & 22 \\ \hline
\end{longtable}

\section{Results}

\subsection{Light Curves}

\begin{figure}
  \begin{center}
   \includegraphics[width=140mm,clip,bb=0 0 360 280]{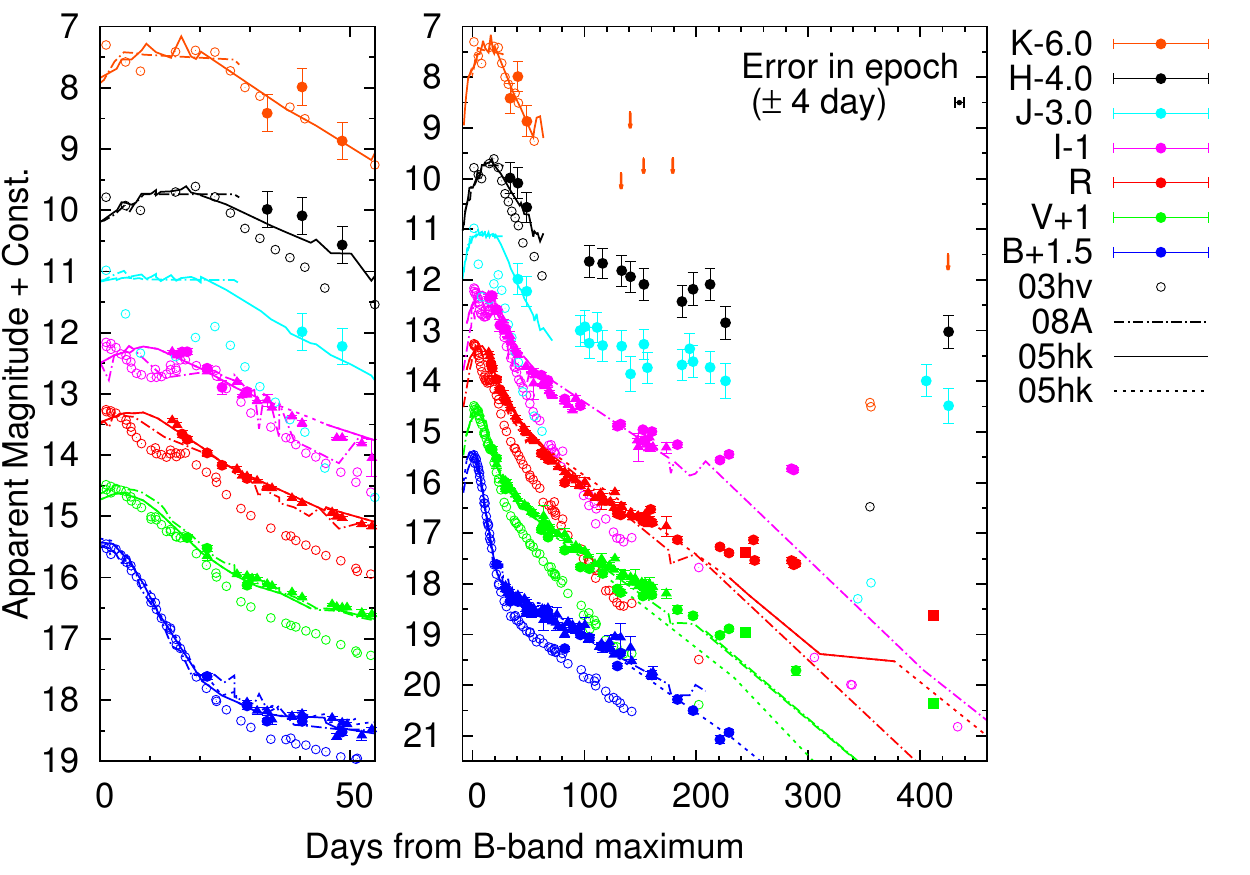}
  \end{center}
  \caption{Multi-band light curves of SN 2014dt.
The filled circles, triangles and squares denote data that 
were obtained at Kanata, OKU, and Subaru, respectively.
The light curve of each band is shifted vertically as 
indicated in the top-right portion of the panel.
We adopted MJD $56950.4 \pm 4.0$ as 0 day, and show its 
uncertainty in the panel (see \S 3.3).
In the left panel, we show the enlarged ones 
for the early phase ($\lesssim$ 60 days).
For comparison, we show the light curves of SN Iax 2005hk 
with solid lines (or dashed ones in cases where separation 
to the neighboring point is $\gtrsim$ 100 days) 
($BVRIJHKs$-bands and $F606W$-band; Phillips et al. 2007, 
Sahu et al. 2008, Friedman et al. 2015, McCully et al. 2010, 
Stritzinger et al. 2015, Krisciunas et al. 2017) and SN 2008A 
($BVRI$, $r'i'$ and $F555W, F622W, F625W, F791W, F775W$-bands) 
with dashed-dotted lines (McCully et al. 2010, Hicken et al. 2012, 
Silverman et al. 2012, Brown et al. 2014, Friedman et al. 2015) 
and SN Ia 2003hv with the open circles (Leloudas et al.2009). 
}\label{14dt_lc}
\end{figure}

Figure \ref{14dt_lc} shows $BVRIJHK_{s}$-band LCs of SN 2014dt.
While we missed the pre-maximum data because SN 2014dt was 
discovered after the maximum, we continued follow-up 
observations for 412 days, which is one of the longest 
multi-band monitorings of SNe Iax.
By comparison with LCs of other SNe Iax, we estimated 
the epoch of the $B$-band maximum as MJD $56950.4 \pm 4.0$ 
(see \S 3.3).
In this paper, we adopted the $B$-band maximum as 0 day.

We estimated the decline rate in the late phase by fitting 
a linear function to the LC.
The decline rate of SN 2014dt is 0.009 $\pm$ 0.001 mag 
day$^{-1}$ between 100 -- 430 days in the $V$-band.
The fitting error includes uncertainties on the phase 
and the photometric error.
Using the data of Singh et al. (2018), we also 
estimate the decline rate up to 200 days in the same way.
The decline rate is 0.008 $\pm$ 0.001 mag day$^{-1}$.
The value is consistent with that obtained from our data.
This is slower than those of 2002cx (0.012 mag day$^{-1}$; 
McCully et al. 2014), 2005hk (0.015 mag day$^{-1}$; 
McCully et al. 2014), 2008A (0.019 mag day$^{-1}$; 
McCully et al. 2014).
These late phase decline rates of SNe Iax except for SN 
2014dt are similar to that of a normal SN Ia SN 2011fe, 
which is 0.0151 mag day$^{-1}$ at 100 -- 300 
days (Zhang et al. 2016).
SN 2014dt shows the slowest decline among SNe Ia and Iax.
Similarly, SN 2014dt evolves much more slowly than normal 
SN 2013hv in NIR.

\begin{figure}
  \begin{center}
   \includegraphics[width=140mm,clip,bb=0 0 360 270]{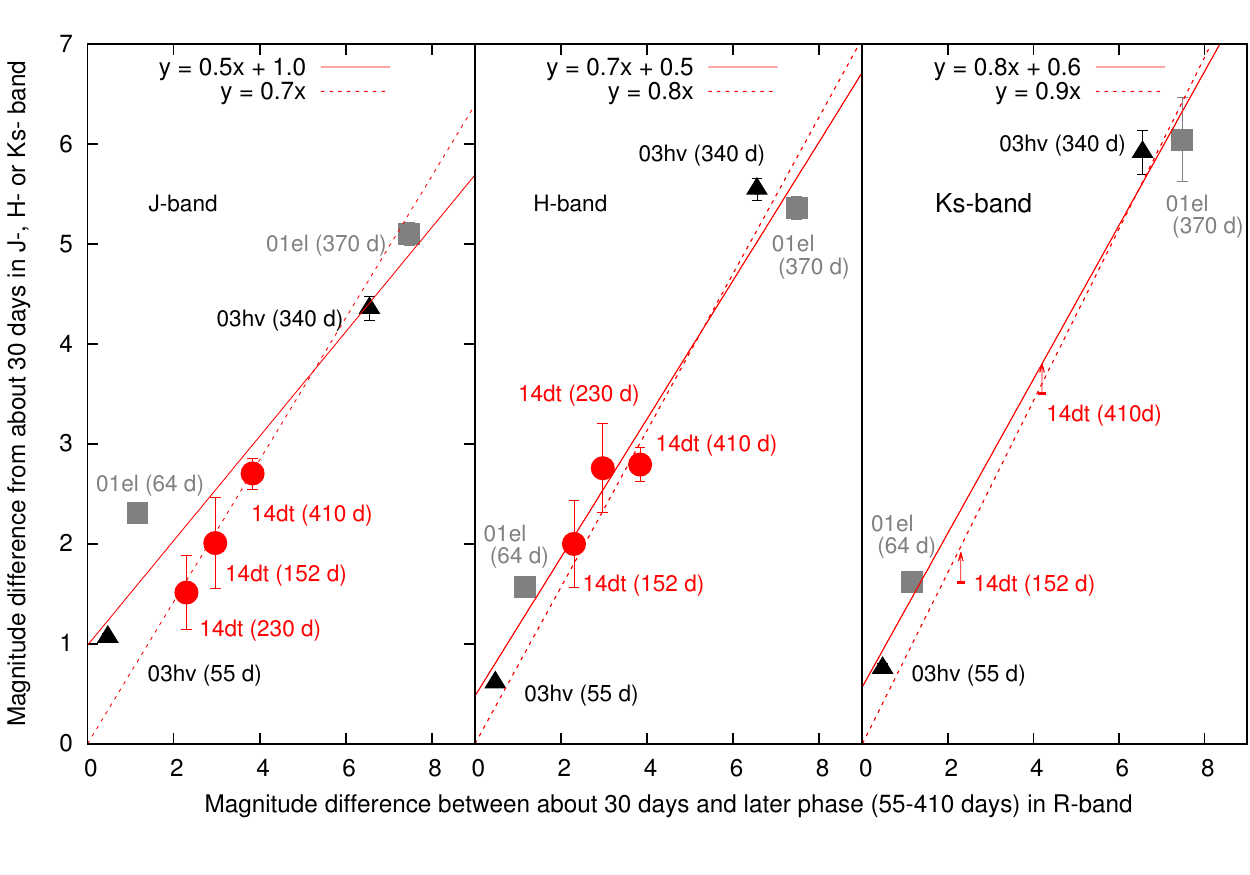}
  \end{center}
   \caption{Relationship between the long-term decline rates in
the optical and NIR bands for SN 2014dt.
The decline rate is measured from $\sim$ 30 days to a given epoch
(152, 230 and 410 days) in the late phase.
The epochs, as measured from the maximum light, are plotted
in parentheses in the figure.
For comparison, we plotted those of normal, non-dust forming
SN Ia 2003hv at 55 and 340 days as black symbols (Leloudas
et al. 2009) and SN 2001el at 64 and 370 days as gray symbols
(Krisciunas et al. 2003, Stritzinger et al. 2007).
The circle, square and triangle symbols denote data obtained
in the $J$, $H$, and $K_{s}$-bands, respectively.
The data points were approximately aligned along straight
lines suggesting that the long-term decline
is homogeneous over optical and NIR bands and that no clear
sign of NIR excess was found in SN 2014dt at $230$--$410$ days
when the MIR excess was observed (see \S 4.1).
}\label{14dt_sed2}
\end{figure}

We compared the long-term decline rates of SN 2014dt at 
different epochs with those of SNe 2001el (Krisciunas 
et al. 2003; Stritzinger et al. 2007) and 2003hv (Leloudas
et al. 2009), which are well-observed, non-dust forming
normal SNe Ia with available late phase optical-NIR
photometric data.
Figure \ref{14dt_sed2} shows the relationship between the
long-term decline rates in optical and NIR bands for
SN 2014dt at three epochs, and for SNe 2001el and 2003hv
at two epochs.
They are well aligned along straight lines,
$m_{\rm NIR}$/$m_{\rm opt} \simeq 0.5$--$0.9$,
suggesting that the declining trends in optical and NIR
wavelengths are similar in those SNe.
This suggests that the slow $V$-band decay traces a slow 
evolution in the bolometric light curves, and it is not 
caused by a color evolution.

\subsection{Spectral Evolution}
\begin{figure}
  \begin{center}
   \includegraphics[width=160mm,clip,bb=0 0 360 250]{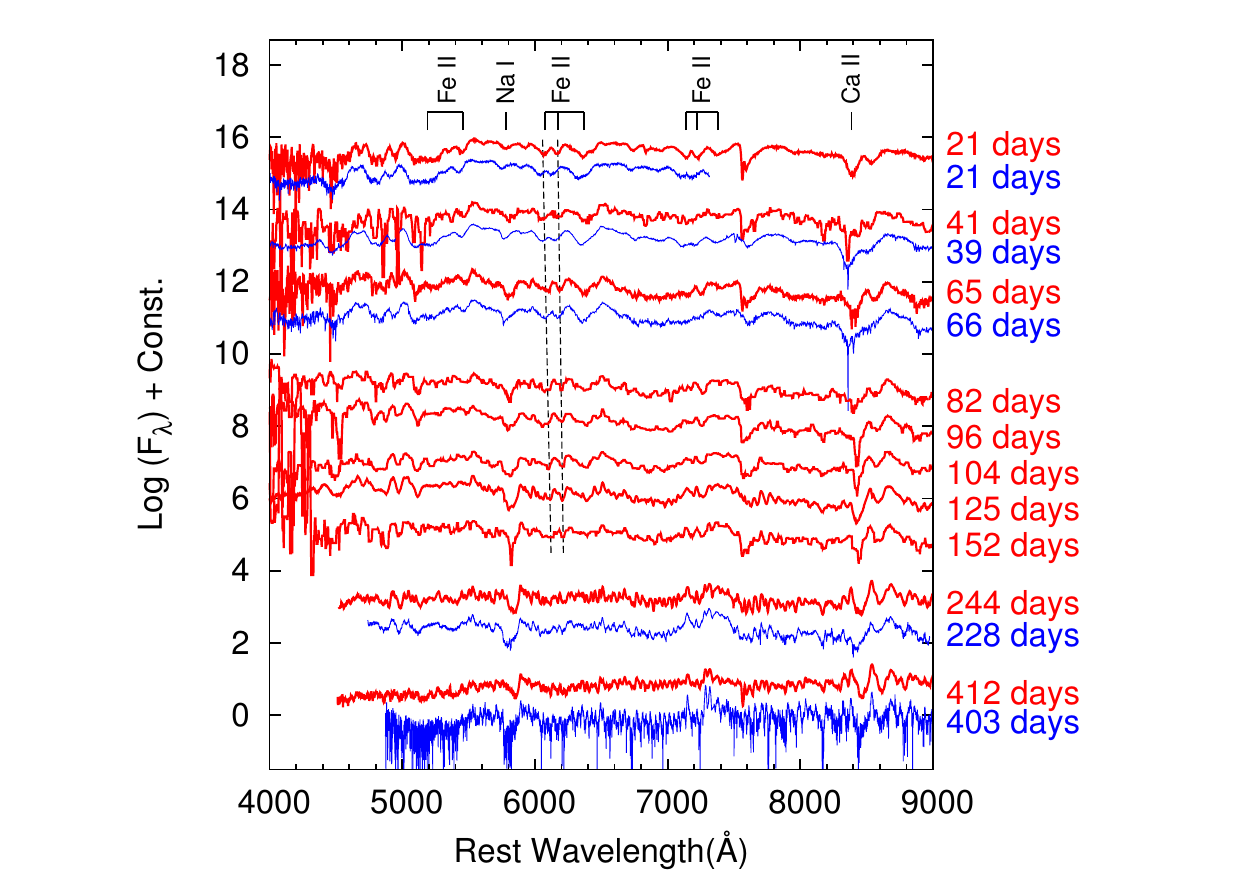}
  \end{center}
   \caption{Spectral evolution of SN 2014dt (red solid lines).
The epoch of each spectrum is indicated on the right outside 
the panel.
The dashed lines show the positions of Fe {\sc ii} 
$\lambda$6149 and $\lambda$6247 lines.
For comparison, we plotted the spectra of SN 2005hk (blue 
lines, Phillips et al. 2007; Blondin et al. 2012; 
Silverman et al. 2012).}\label{14dt_sp}
\end{figure}

\begin{figure}
  \begin{center}
   \includegraphics[width=120mm,clip,bb=0 0 360 270]{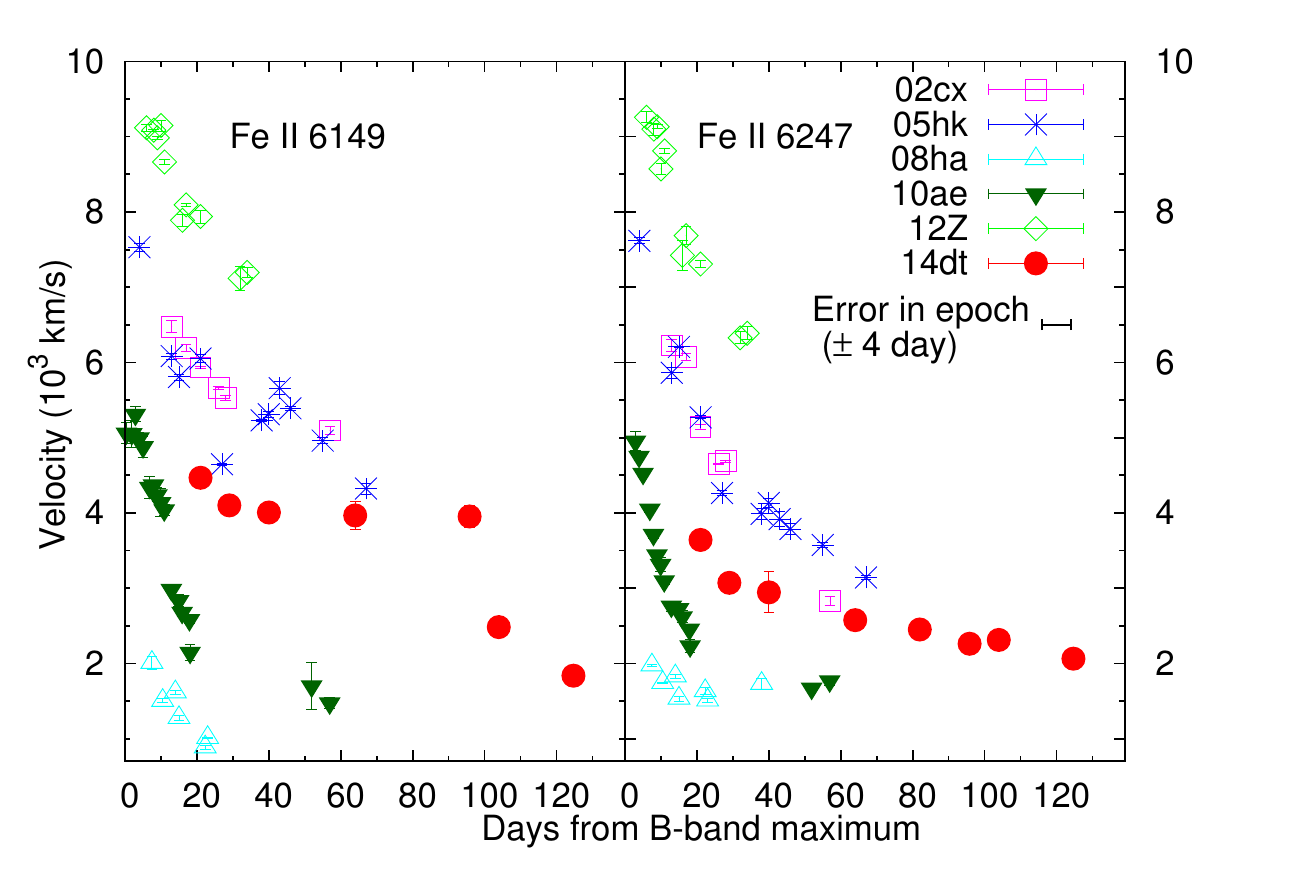}
  \end{center}
   \caption{Line velocities of Fe {\sc ii} $\lambda$6149 
and $\lambda$6247 in SN 2014dt.
For comparison, we plotted those in SNe Iax 2002cx (Li et al. 2003), 
2005hk (Phillips et al. 2007; Blondin et al. 2012; Silverman et al. 
2012), 2008ha (Foley et al. 2009; Valenti et al. 2009), 2010ae 
(Stritzinger et al. 2014) and 2012Z (Stritzinger et al. 2015; 
Yamanaka et al. 2015).
We measured the velocity of the comparison SNe in the same way as 
for SN 2014dt.}\label{14dt_vel}
\end{figure}

Figure \ref{14dt_sp} shows optical spectra of SN 2014dt 
from 21 days through 412 days.
The early phase spectra are characterized by absorption 
lines of Na {\sc i D}, Fe {\sc ii}, Co {\sc ii}, and the 
Ca {\sc ii} IR triplet (Branch et al. 2004; Jha et al. 2006; 
Sahu et al. 2008).
The line identification is partly in debate due to 
non-negligible line blending (see below).
The line of Si {\sc ii} $\lambda$6355 is likely hidden 
by that of Fe {\sc ii} $\lambda$6456 at 21 days.
These lines are usually seen in SNe Iax at similar epochs.
The spectral features of SN 2014dt at $21$--$64$ days 
closely resemble those of SN 2005hk at similar epochs, 
except that the line width (and the blueshift) of each 
absorption feature in SN 2014dt is slightly narrower 
(and smaller) than those in SN 2005hk.
This suggests that SN 2014dt had a smaller expansion 
velocity in the early phase.
In Figure \ref{14dt_vel}, we compare the line velocities 
of Fe {\sc ii} $\lambda$6149 and $\lambda$6247 lines in 
SN 2014dt with those in other SNe Iax.
We measured the velocities by a Gaussian fit to each 
absorption line.
In each spectrum, we measured it multiple times for each lines.
The scatter of the line velocities due to the measurements 
$\sim$400 km s$^{-1}$.
The derived velocity was roughly 4000 km s$^{-1}$ at $21$ 
days and decreased to $2000$--$3000$ km s$^{-1}$ at 
$100$--$120$ days. 
This is slower than those of SNe 2002cx, 2005hk and 2012Z, 
whereas it is slightly faster than those of SNe 2009ku 
(Narayan et al. 2011) and 2014ck (Tomasella et al. 2016).

\begin{figure}
  \begin{center}
   \includegraphics[width=160mm,clip,bb=0 0 360 250]{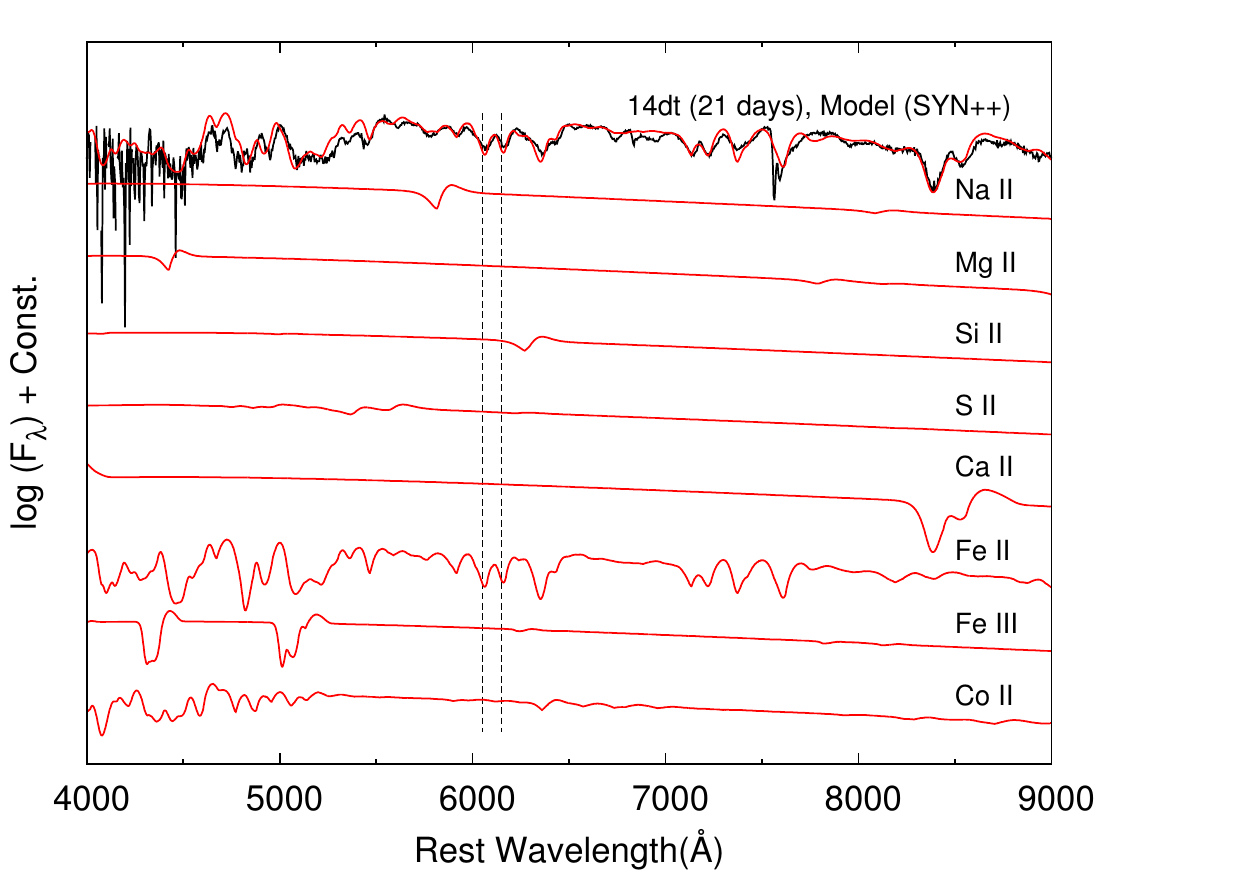}
  \end{center}
   \caption{Comparison of the observed spectrum for SN 2014dt 
at 21 days (black line) and the synthesized spectra calculated 
with {\it SYN++} code (red lines).
We also show the synthesized spectra of each species indicated 
in the panel. 
The dashed lines show the positions of Fe {\sc ii} $\lambda$6149 
and $\lambda$6247 lines.
It is likely that Fe {\sc ii} $\lambda$6149 and $\lambda$6247 
are little blended with other species (see text).
}\label{sp_mod}
\end{figure}

Here, we should be careful in the treatment of line 
velocities, because the spectra of SNe Iax show many 
intermediate-mass and Fe-group elements, and it is thus 
difficult to identify individual lines (e.g., Szalai 
et al. 2015) even with the less line blending due to 
the slower expansion than normal SNe Ia.
We synthesized model spectra using {\it SYN++} code 
(Thomas et al. 2011) to provide reliable line identifications.
The synthetic spectrum explained the overall features 
well in the spectrum (Figure \ref{sp_mod}).
At around 6000 \AA\ , the spectral lines of Fe {\sc ii} 
likely dominate the observed features, and Fe {\sc ii} 
$\lambda$6149 and $\lambda$6247 suffered from little 
contamination by other lines (e.g., Co {\sc ii}) as 
suggested in previous studies (i.e., Stritzinger et al. 
2014 for SN 2010ae and Szalai et al. 2015 for SN 2011ay).
This supports the idea that the slower expansion of 
SN 2014dt seen in Figure \ref{14dt_vel} is real in nature.

In the late phase, 244 and 412 days, the spectra are 
characterized by many narrow emission lines.
Many permitted lines, such as Fe {\sc ii}, and Ca {\sc ii}, 
are seen in addition to forbidden lines, such as 
[Fe {\sc i}] $\lambda$7155 and [Ca {\sc ii}] 
$\lambda\lambda$7291, 7324.
The spectral features at 241 days are still quite similar 
to those of SN 2005hk at 232 days except for some minor 
differences, such as the line ratios among emission lines 
around $8400$--$8700$ \AA \ and the width and blueshift of 
the Na {\sc i} D absorption line.
Generally, the strength ratio of a forbidden line to a 
permitted one is an index of the density of the 
line-emitting region; a smaller ratio indicates a larger 
density (e.g., Li \& McCray 1993).
We derived the strength of the emission lines simply by a 
single-Gaussian fit to each line profile.
In the spectrum of SN 2014dt at 244 days, the ratio of 
[Ca {\sc ii}] $\lambda\lambda$7291, 7324 to Ca {\sc ii} 
$\lambda$8542 was 0.7 $\pm$ 0.1, whereas that in the 
spectrum of SN 2005hk at 228 days was much larger, 
2.5 $\pm$ 0.1.
This suggests that the density of the line emitting 
region in SN 2014dt is higher than that in SN 2005hk.

The spectral evolution through $\simeq 410$ days is 
markedly different between SNe 2014dt and 2005hk.
SN 2014dt showed no significant change except that some 
Fe {\sc ii} lines at the shorter wavelengths become weaker.
However, in SN 2005hk, the forbidden lines ([Fe {\sc i}] 
$\lambda$7155 and [Ca {\sc ii}] $\lambda\lambda$7291, 7324) 
become stronger relatively to the continuum and most 
emission lines become narrower.
These indicate that the spectral evolution of SN 2014dt 
in the late phase was slow, which is likely consistent 
with the slow decline of the optical and NIR luminosity.

\subsection{Estimation of the Maximum Epoch}

It is useful to restrict the epoch of maximum brightness 
even roughly for comparison with other SNe and theoretical 
models.
We emphasize that the main arguments in this paper are 
based on the long-term evolution through the late phase, 
and thus uncertainty in the maximum epoch and explosion 
date do not alter our main conclusions. 
However, some of our additional arguments 
depend on the epoch of maximum brightness. 
In this subsection, we describe how we estimated the 
maximum epoch from the observational data of SN 2014dt, 
for which the maximum date is indeed missing.

\begin{figure*}
  \begin{center}
   \includegraphics[width=120mm,clip,bb=0 0 360 270]{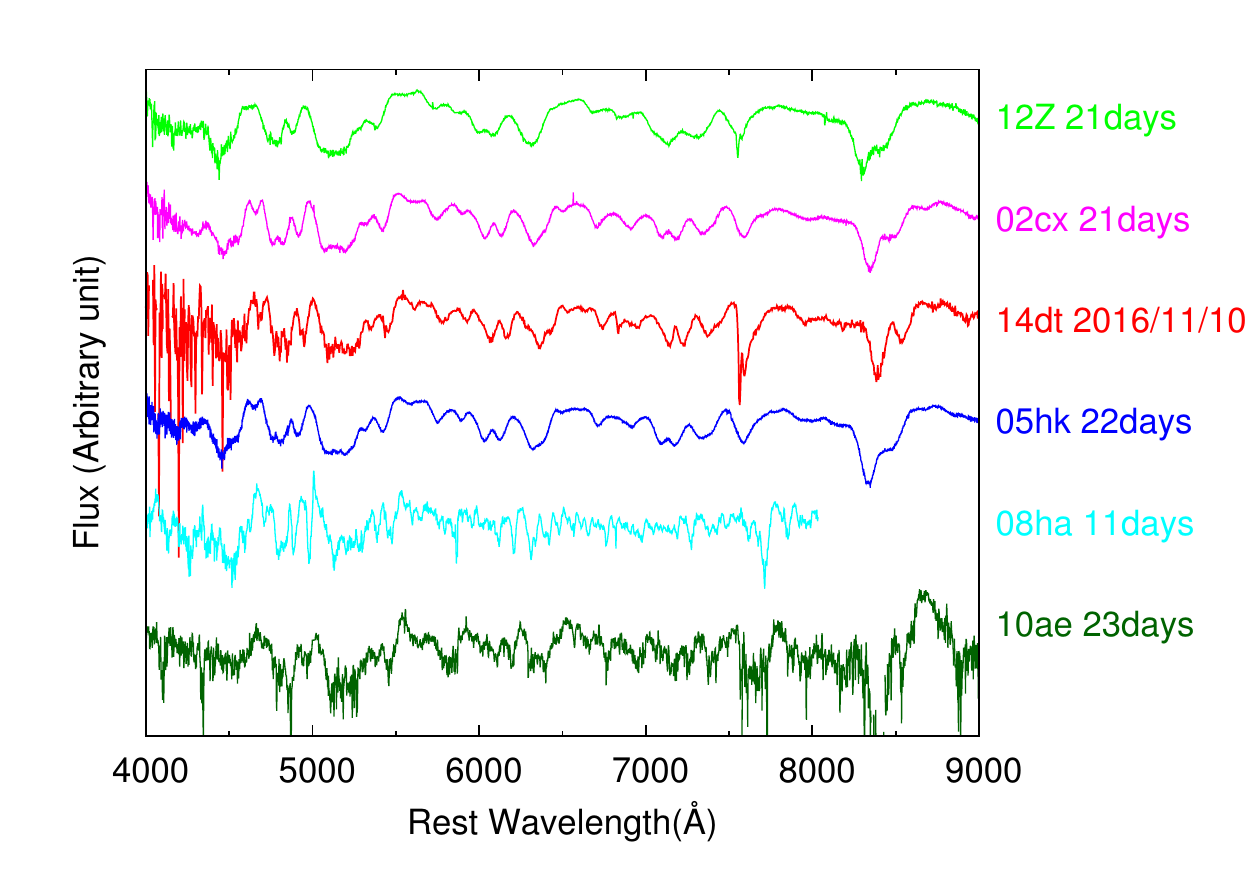}
  \end{center}
  \caption{Spectral comparison of SN 2014dt on 2016 November 
10 with other SNe Iax; 2012Z (Stritzinger 
et al. 2015), 2002cx (Li et al. 2003), 2005hk (Blondin et al. 
2012), 2008ha (Foley et al. 2009), and 2010ae (Stritzinger 
et al. 2014). 
The phase of each spectrum is $\sim$20 days except for 
SN 2008ha for which no spectrum at $\sim$ 20 days was available.}\label{comp_sp}
\end{figure*}

Figure \ref{comp_sp} shows a spectral comparison of 
SN 2014dt and other SNe Iax.
The spectrum of SN 2014dt on November 10 shows a close 
resemblance to those of SNe 2002cx at 21 days and 
2005hk at 22 days, SN 2012Z at 21 days.
The absorption lines of SNe 2008ha and 2010ae are 
much narrower; however, the overall spectral features 
are still similar to those of other SNe Iax at similar 
epochs.
This spectral comparison could provide a tentative 
estimation of the epoch of SN 2014dt: 2014 November 10 
was around $21$--$22$ days after the maximum.

Next, we considered an estimation of the epoch based on 
the LCs.
Indeed, the multi-band LCs of SN 2014dt apparently closely 
followed those of SNe 2002cx, 2005hk, and 2012Z in the 
post-maximum phase, until $\sim 80$ days after the explosion.
In this period, the extinction-corrected colors and 
the spectra of SN 2014dt also resembled those of SNe 
2002cx, 2005hk and 2012Z except for a slightly slower 
expansion velocity (see \S 3.2).
These data allowed us to conclude that the multi-band LCs 
of SN 2014dt even before the discovery were similar 
to those of SNe 2002cx, 2005hk and 2012Z.
To further quantify the similarity and provide a reasonable 
estimate on evolution of the LCs of SN 2014dt before the 
discovery, we examined whether and how the multi-band LCs 
of these template SNe matched to those of SN 2014dt. 
The details are given in the Appendix. 
Here we provide a summary of our analyses.

First, the initial guess in the phase was adopted so that it 
was 21 days after the maximum on 2014 November 10. 
For a template LC of each SN in each bandpass, we allowed the 
shifts along the time axis $\Delta t$ and the magnitude axis 
$\Delta M$ within given ranges ($| \Delta t | \leq 10$ days 
and $| \Delta M | \leq 0.3$ mag). 
For each set of $\Delta t$ and $\Delta M$, we computed a 
residual between the SN 2014dt and the hypothesized 
template LC between the discovery date of SN 2014dt and 
60 days after the discovery. 
Then, we estimated the plausible ranges of $\Delta t$ and 
$\Delta M$ requiring that the residual should be smaller 
than a given specific value.
Performing the same procedure independently for $BVRI$-band 
LCs, we determined the allowed range of $\Delta t$ and 
$\Delta M$ so that the fits in all of the bands were mutually 
consistent. 
However, we found that the multi-band LCs of SN 2002cx did 
not provide a consistent solution among different bands; thus 
we concluded that the LCs of SN 2002cx should be omitted from 
the template LCs. 
As a result, we had a series of `combined' LCs for SN 2014dt, 
in which the pre-discovery part was replaced by the LC of 
either SN 2005hk or SN 2012Z, taking into account the 
uncertainty in $\Delta t$ and $\Delta M$. 
Further details can be found in Appendix.
Thus, we derived the epoch of $B$-band maximum as MJD 
$56950.4 \pm 4.0$ (2014 October 20.4 UT), which was in 
accordance with the results of the spectral comparison and 
the also with the previous estimate (Foley et al. 2016, 
Fox et al. 2016 and Singh et al. 2018).

\subsection{Maximum Magnitude v.s. Decline Rates}

For the `combined' LCs constructed in \S 3.3 (see also 
Appendix), we estimated the maximum magnitude and the 
decline rate, $\Delta m_{15}$, as well as the epoch of 
the maximum light in each of the $VRI$-bands (Table 5).
After correcting the distance modulus and the total 
extinction (see \S 1), we derived the absolute maximum 
magnitude in the $B$-band, $M_{B,{\rm max}} =$ $-16.82$--$-17.34$ 
mag\footnotemark[10]\footnotetext[10]{
$M_{B,{\rm max}} =$ $-15.31$--$-17.27$ mag using the distance 
obtained the CO Tully-Fisher relation, $M_{B,{\rm max}} =$ 
$-15.84$--$-17.12$ mag using the distance obtained the 
H{\sc i} Tully-Fisher relation.}.
It is $\sim$1 mag fainter than that obtained by Singh 
et al. (2018), which is apparently originated from the 
systematic difference in the $B$-band magnitude of the 
companion stars (see \S 2) and in the adopted distance 
modulus.
Our $M_{B, {\rm max}}$ was slightly ($0.2$--$1.2$ mag) 
fainter than those of SNe Iax 2002cx 
($M_{B,{\rm max}} = -17.55 \pm 0.34$ mag; Li et al. 2003), 
2005hk ($M_{B,{\rm max}} = -18.02 \pm 0.32$ mag; Phillips 
et al. 2007, $M_{B,{\rm max}} = -18.00 \pm 0.25$ mag; 
Stritzinger et al. 2015), and 2012Z ($M_{B,{\rm max}} = -18.27 \pm 0.09$ 
mag; Stritzinger et al. 2015, $M_{B,{\rm max}} \sim -17.61$ mag; 
Yamanaka et al. 2015 \footnotemark[11]\footnotetext[11]{The 
difference in the reported absolute magnitudes is due to the 
different values of the distance modulus and the dust 
extinction adopted in the two studies. 
Stritzinger et al. (2015) adopted $\mu = 32.59 \pm 0.019$ mag 
and $E(B-V) = 0.11 \pm 0.03$ mag. On the one hand, Yamanaka et al. 
(2015) adopted $\mu = 32.4 \pm 0.3$ mag and $E(B-V) = 0.036$ mag.}), 
whereas it was $3.1$--$3.6$ mag brighter than that of SN 2008ha 
($M_{B,{\rm max}} = -13.74 \pm 0.15$ mag; Foley et. al. 2009, 
$M_{B,{\rm max}} = -13.79 \pm 0.14$ mag; Stritzinger et al. 2014).
It is noted that this estimate conservatively included the 
uncertainty in the `combined' LCs in terms of the template LCs 
(SN 2005hk or SN 2012Z) and the fitting uncertainties in 
$\Delta t$ and $\Delta M$.

\begin{longtable}{lccc}
\caption{Peak magnitude and its epoch of SN 2014dt in each photometric band \footnotemark[12]}
\hline
Band & Maximum date (MJD) & Maximum magnitude & $\Delta m_{15}$ \footnotemark[13]\\
\hline
\endhead
\hline
\endfoot
\hline
\multicolumn{4}{l}{\hbox to 0pt{\parbox{180mm}{\footnotesize
\footnotemark[12] They are deduced from the light curves
combined with those of SNe 2005hk \newline and 2012Z 
since the observed light curves of SN 2014dt do not cover 
the \newline peak magnitude. See \S 3.3 for details.
\par\noindent
\footnotemark[13] Magnitude difference between 0 days and
15 days after the maximum.
}\hss}}
\endlastfoot
   $B$ & 56950.4 $\pm$ 4.0 & $13.88$--$13.76$ & $1.43$--$1.57$\\
   $V$ & 56954.9 $\pm$ 4.0 & $13.56$--$13.50$ & $0.73$--$0.91$\\
   $R$ & 56957.5 $\pm$ 4.0 & $13.47$--$13.28$ & $0.51$--$0.66$\\
   $I$ & 56958.5 $\pm$ 4.0 & $13.63$--$13.30$ & $0.48$--$0.66$\\ \hline
\end{longtable}

\begin{figure}
  \begin{center}
   \includegraphics[width=120mm,clip,bb=0 0 360 270]{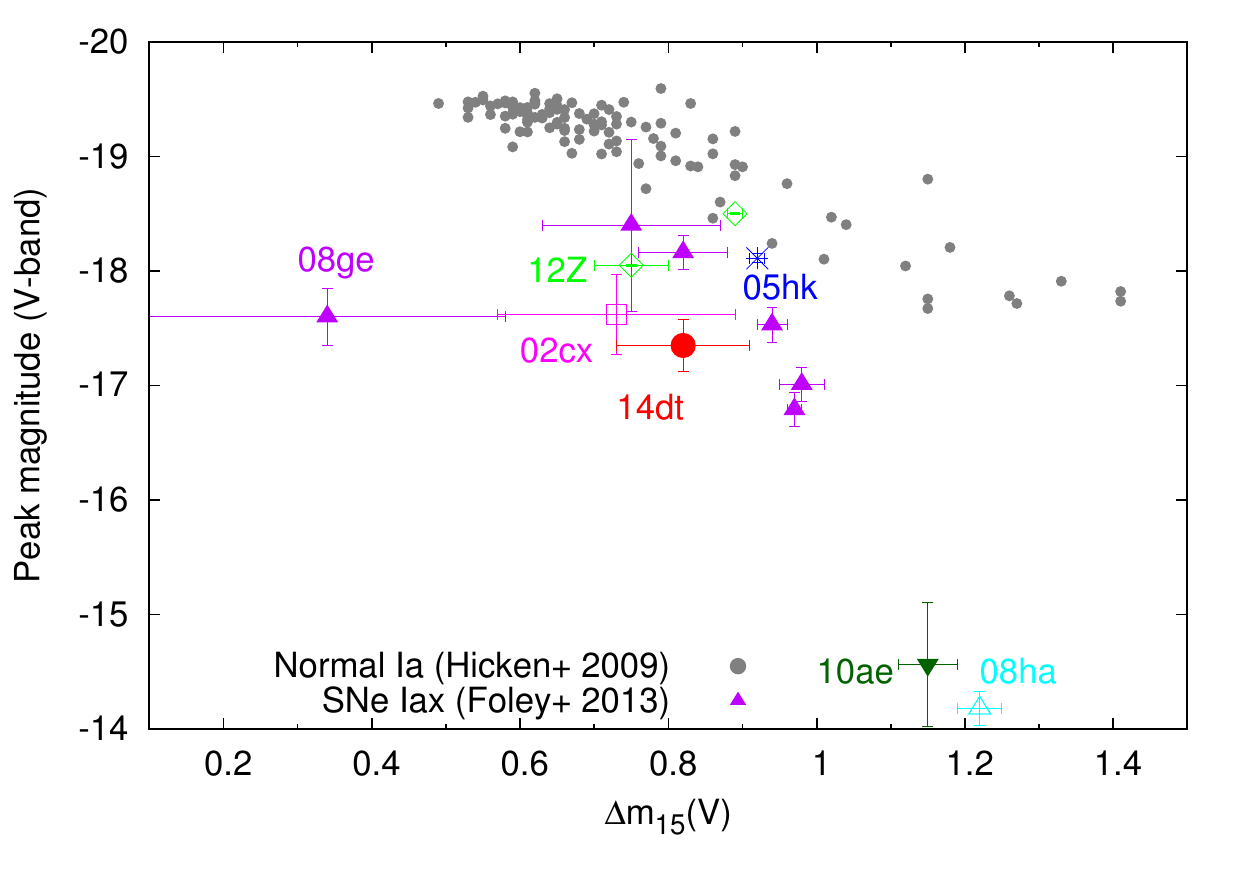}
  \end{center}
   \caption{Correlation of the peak absolute magnitudes 
in the $V$-band with the decline rate $\Delta m_{15}(V)$.
The gray filled circles are the data points of normal 
SNe Ia (Hicken et al. 2009), and the other symbols are for 
SNe Iax (Li et al. 2003; Foley et al. 2009; Foley et al. 2013; 
Stritzinger et al. 2014; Stritzinger et al. 2015; 
Yamanaka et al. 2015).
For some SNe Iax, the SN IDs are indicated one by one.
}\label{decline}
\end{figure}

\begin{figure}
  \begin{center}
   \includegraphics[width=120mm,clip,bb=0 0 360 270]{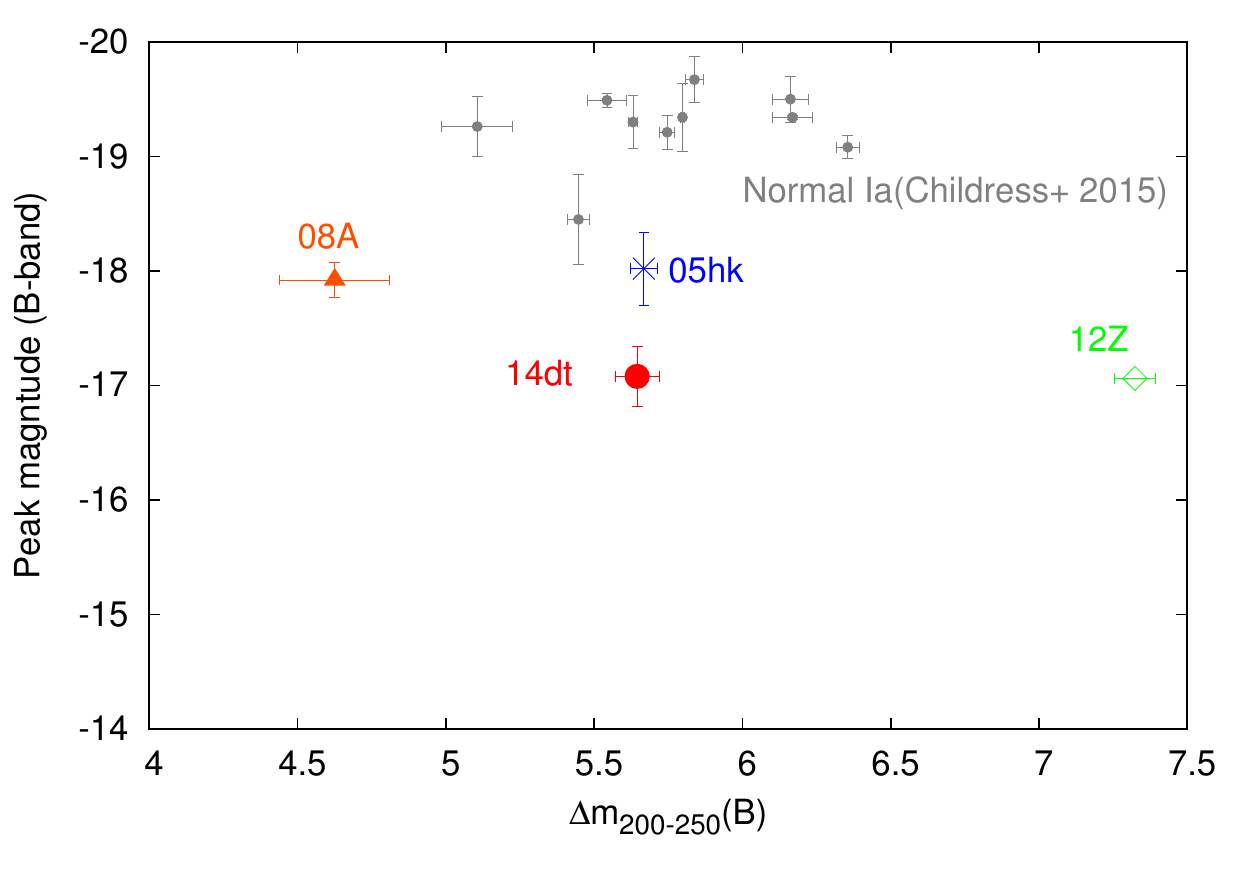}
  \end{center}
   \caption{Relationship between the peak absolute 
magnitudes in the $B$-band and the long-term decline rate; 
that is, the magnitude difference between $0$ and $200$--$250$ 
days after the maximum light (see \S 3.4), 
in the $B$-band.
The data for normal SNe Ia are from Childress et al. (2015).
}\label{decline2}
\end{figure}

The derived decline rate of SN 2014dt was 
$\Delta m_{15}(B) = 1.43$--$1.57$ mag.
As shown in Table 5, the decline rate was smaller at 
longer wavelengths.
Figure \ref{decline} shows the relationship between 
$M_{V, {\rm max}}$ and $\Delta m_{15}(V)$ in some SNe Iax 
and normal SNe Ia.
SN 2014dt apparently belonged to the group of the bright 
SNe Iax including SNe 2002cx, 2005hk and 2012Z.
From 60 through 410 days, the differences in LCs from those 
of SN 2005hk became significant (Figure \ref{14dt_lc}); 
SN 2014dt showed a slower decline than SN 2005hk.
This is confirmed in Figure \ref{decline2} where the 
`long-term decline rates' ($0$ to $200$--$250$ days) in the 
$B$-band were plotted.
Among three SNe Iax for which $B$-band photometric data at 
$200$--$250$ days existed SN 2014dt showed the smallest 
long-term decline rate, and was the smallest compared to 
larger samples of normal SNe Ia.
We found no clear trend between the absolute maximum 
magnitude and the long-term decline rate among SNe Iax 
samples or normal SNe Ia samples (see \S4. for discussion).
The large error in $\Delta m_{200-250}(B)$ stemmed from our 
conservative error estimated in the maximum magnitude. 
Indeed, it is clear from Figure \ref{14dt_lc} that the LCs 
of SN 2014dt were much flatter in the late phase than 
those of SN 2005hk (see \S3.1).

\subsection{Spectral Energy Distribution}

Our observations may provide the first opportunity to 
examine the long-term, optical and NIR multi-band photometry 
for SNe Iax, which enables studying the evolution of SED 
through the late phase, including any possible contribution 
from circumstellar (CS) dust.
At longer wavelengths, Fox et al. (2016) reported that 
SN 2014dt showed brightening around 300 days at mid-infrared 
(MIR) wavelengths and suggested that thermal emissions from 
pre-existing dust grains could cause MIR excess.

\begin{figure}
  \begin{center}
   \includegraphics[width=120mm,clip,bb=0 0 360 270]{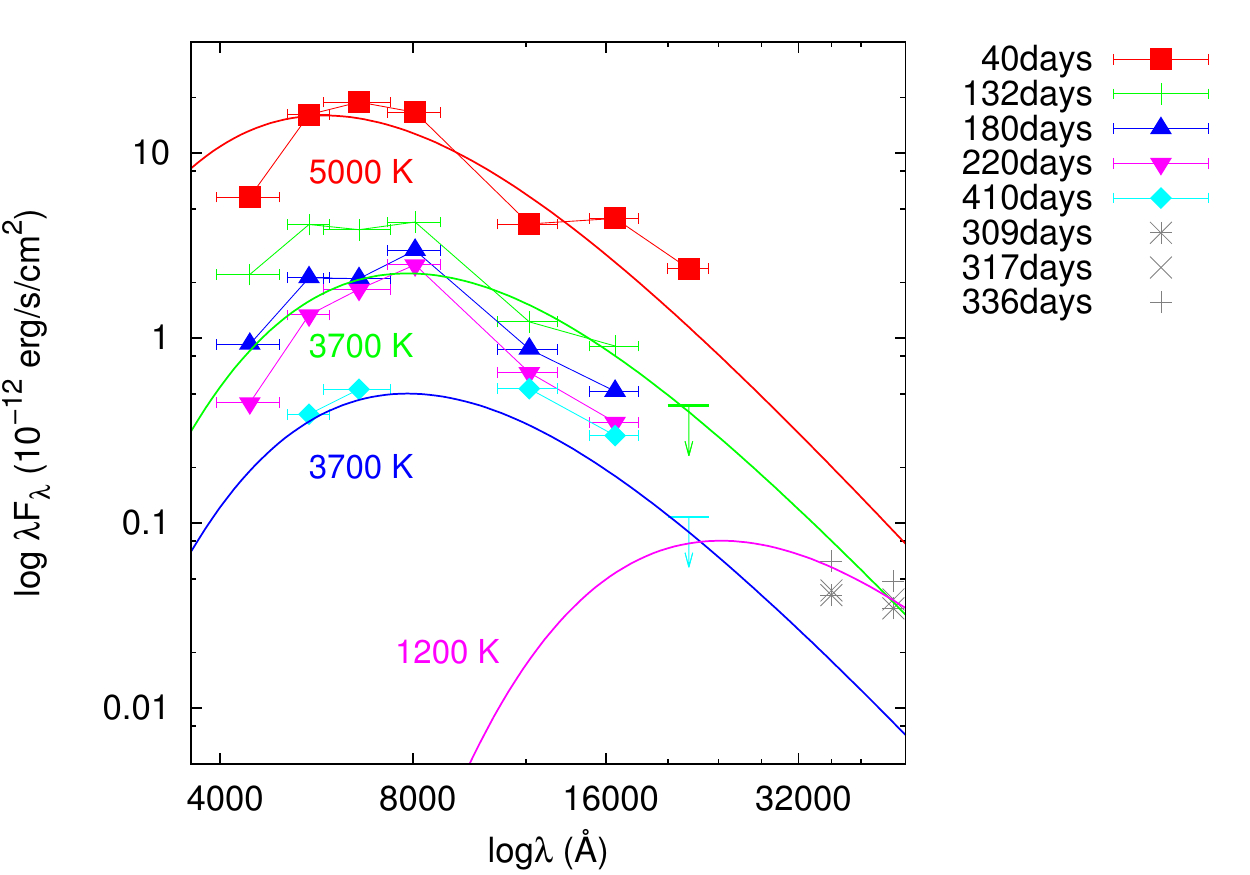}
  \end{center}
   \caption{The evolution of the SED of SN 2014dt. 
For data at 180, 220, and 410 days, we plotted the combined 
data at similar epochs (difference $\lesssim$ 14 days) 
because we did not always have the data exactly at the 
same epoch.
The data at mid-infrared wavelengths (the 
gray points) are from Fox et al. (2016).
For comparison, we plotted the blackbody spectra for 5000, 
3700, and 1200 K by the solid lines.
}\label{14dt_sed}
\end{figure}

We derived SEDs of SN 2014dt by combining the optical and 
NIR photometric data at similar epochs (difference $\lesssim$ 
14 days), and plotted them in Figure \ref{14dt_sed} together 
with representative blackbody (BB) spectra.
The BB temperature is roughly an indicator of the photospheric 
temperature.
Initially (between 40 days and 132 days) the BB temperature 
seemed to decrease rapidly from $\sim5000$ K to $\sim3700$ K, 
but thereafter the BB temperature showed no significant change 
through 410 days, 
whereas the peak flux gradually decreased.
For some SNe Iax, McCully et al. (2014) points out that 
temperature evolution obtained by the ratio of the permitted 
Ca {\sc ii} lines is slower than $t^{-1}$ or even constant. 
This might be commonly seen in SNe Iax.
However, the fluence ($\lambda F_{\lambda}$) at MIR wavelengths 
gradually increased during the period of $309$--$336$ days 
(Fox et al. 2016), although it was still much smaller 
than those in optical and NIR wavelengths.
The spectral slope of the MIR excess was inconsistent with 
the main SED component peak at optical wavelengths, 
and thus the MIR excess would only represent a minor, 
additional component. 

\section{Discussion}
\subsection{Interpretation of the SED Evolution}

As shown in Figure \ref{14dt_sed}, the main SED component 
($\sim 3700$ K BB emission) dominated the optical and 
NIR flux from 180 days to 410 days.
This long-lasting SED component could be an analog to the 
component that Foley et al. (2016) found for SN 2005hk by 
continuum fits to late phase spectra.
The authors suggested that the long-lasting photospheric 
emission could have originated in a wind launched from a 
bound remnant in a context of the weak deflagration model.
It is plausible that this photospheric emission component
corresponded to the main SED component derived for SN 2014dt.

The combination of the bolometric luminosity and the temperature 
obtained through the SED fitting suggests that the photosphere is  
located at $R_{\rm ph} \sim 6.2 \times 10^{14}$ cm at 230 days. 
Assuming that the bolometric luminosity follows the $^{56}$Co decay, 
then the nearly unchanged temperature suggests 
$R_{\rm ph} \sim 2.9 \times 10^{14}$ cm at 410 days, i.e., the 
photospheric radius has decreased by about a factor of two from 
230 days to 410 days. 
Assuming that the opacity is $0.1$ cm$^{2}$ g$^{-1}$, then 
$R_{\rm ph}$ (230 days) suggests that 
$\sim 8 \times 10^{-3} M_\odot$ of the material is confined 
within this radius. 
{\it If} the photosphere could be located within the homologously 
expanding inner ejecta, then the corresponding velocity is 
$\sim 300$ km s$^{-1}$ and $\sim 80$ km s$^{-1}$ at 230 and 410 
days, respectively. 
To compare to these values of the velocity, the late phase spectra 
at 232 and 412 days show some permitted lines whose widths are 
marginally resolved by our spectroscopic observations. 
Given the resolution of 650, the velocities of these permitted 
lines should be an order of 100 km s$^{-1}$, while the precise 
measurement is not possible. 
In the discussion below, we just use the rough value (a few 100 
km s$^{-1}$) in a qualitative way.

The photospheric radius far exceeds the size of a red supergiant, 
and therefore it is unlikely that this is created by a (nearly 
hydrostatic) atmosphere of a hot (remnant) white dwarf within the 
weak deflagration scenario. 
The overall feature may be consistent with a scenario that the 
photosphere is located deep in the homologously expanding ejecta, 
as the inferred velocity is consistent with those measured in the 
spectra but within the spectral resolution. 
Indeed, Foley et al. (2016) argued against this interpretation 
the photosphere in the ejecta for SN 2005hk, 
by pointing out the discrepancy (in a snapshot spectrum) 
in the velocities measured in these two different approach. 
If the photosphere is in the homologously expanding ejecta, 
it predicts that the associated line velocity 
decreases as time goes by (see above, i.e., by a factor of nearly 
4 from 200 to 400 days), and thus an accurate measurement of the 
line velocities {\it and their evolution} is a key in testing this 
scenario with the photosphere located deep in the 
ejecta. 
This scenario on the other hand naturally explains the decrease in 
the photospheric radius, as the density of the emitting material 
decreases. 

Alternatively, the photosphere might have formed within a wind 
lunched by the possible bound remnant white dwarf, as suggested by 
Foley et al. (2016) for SN 2005hk. 
In this case, a line velocity has no direct link to the photospheric 
radius but determined by the velocity of the wind. 
Still, it should be an order of at least $100$ km s$^{-1}$ to create 
the photosphere at $R > 10^{14}$ cm. 
The mass within the photosphere ($\sim 8 \times 10^{-3} M_\odot$ as 
measured at 230 days) is translated to the wind mass loss rate of 
$\sim 10^{-2} M_{\odot}$ yr$^{-1}$, and the kinetic power of 
$\sim 3 \times 10^{37}$ erg s$^{-1}$ to $10^{39}$ erg s$^{-1}$ 
for the wind velocity of $\sim 100 - 1000$ km s$^{-1}$. 
This is roughly the Eddington luminosity for the expected bound 
remnant, thus is in line with the wind scenario. 
A possible drawback of the wind scenario is that there is no obvious 
reason about why the photosphere radius has been decreased, as the 
`stationary wind' would not provide a change in the density scale 
to change the position of the photosphere. 

In summary, it is not possible to make a strong conclusion on the 
origin of the photospheric component. 
Both of the ejecta model and the wind model have advantage and 
disadvantage. 
In any case, we add several new constraints on the nature of this 
emission to the suggestion by Foley et al. (2016); for example, 
any model should explain the recession of the photosphere, and 
the mass of the material must be at least 
$\sim 10^{-3} M_{\odot}$. 
In either case, the need of the high-density material in the 
inner part of the ejecta could support the weak deflagration 
scenario and the existence of the bound remnant. 

\subsection{Interpretation of the MIR Excess}
Here we briefly discuss the origin of the minor MIR excess 
(\S 3.5).
Foley et al. (2016) suggested that the emissions from 
circumstellar dust seemed unlikely, because neither 
additional reddening nor narrow absorption lines due to a 
possible circumstellar interaction has been observed in 
late phase optical spectra of SNe Iax.
They further argued against thermal emissions from newly 
formed dust grains within the ejecta, based on the lack of 
expected signatures such as the blueward shift of the 
spectral lines and additional reddening. 
Alternatively, the authors suggested that the emissions 
from a bound remnant with a super-Eddington wind were 
plausible as a source of the MIR excess.
However, as shown above, the main component that we derived 
for SN 2014dt was indeed an analog to what 
Foley et al. (2016) suggested for SN 2005hk, representing 
a possible bound remnant.
By this component alone, it is hard to explain the MIR 
excess.

Thus, we speculate that an echo by circumstellar material 
(CSM) is the most likely interpretation of the MIR excess.
This would not contradict the lack of strong extinction and 
narrow emission lines.
For example, the CSM of the over-luminous SN Ia 2012dn 
is likely to be located off the line of sight, and the 
separation between the CSM and the SN is too far to allow 
a strong interaction within the first few years after the 
explosion (Yamanaka et al. 2016, Nagao et al. 2017).
A similar configuration may apply to the case of SN 2014dt.
 
As mentioned in \S 3.5, the MIR excess only provided a 
minor contribution to the SED.
However, it may not be negligible in the $K_{s}$-band at 
410 days if the continuum from a hot dust ($\sim 1200$ K) 
is the origin of the MIR excess and it may contaminate the 
NIR fluxes (see Figure \ref{14dt_sed}).
Thus, it is interesting to check for possible contamination 
in our NIR data.
If the possible additional component to the intrinsic 
SN emission in the NIR bands exists, the data point should 
be significantly lower than the straight lines in Figure 
\ref{14dt_sed2} (see \S 3.1).
Thus, we exclude the existence of the additional 
component in the NIR bands, and the dust should not be hot 
(i.e., $\lesssim 1200$ K) if thermal emissions from dust is 
the origin of the MIR excess.

%\begin{figure}
%  \begin{center}
%   \includegraphics[width=120mm,clip,bb=0 0 360 270]{figure8.eps}
%  \end{center}
%   \caption{Relationship between the long-term decline rates in 
%the optical and NIR bands for SN 2014dt.
%The decline rate is measured from $\sim$ 30 days to a given epoch 
%(152, 230 and 410 days) in the late phase.
%The epochs, as measured from the maximum light, are plotted 
%in parentheses in the figure.
%For comparison, we plotted those of normal, non-dust forming 
%SN Ia 2003hv at 55 and 340 days as black symbols (Leloudas 
%et al. 2009) and SN 2001el at 64 and 370 days as gray symbols 
%(Krisciunas et al. 2003, Stritzinger et al. 2007).
%The circle, square and triangle symbols denote data obtained 
%in the $J$, $H$, and $K_{s}$-bands, respectively.
%The data points were approximately aligned along straight 
%lines suggesting that the long-term decline 
%is homogeneous over optical and NIR bands and that no clear 
%sign of NIR excess was found in SN 2014dt at $230$--$410$ days 
%when the MIR excess was observed.}\label{14dt_sed2}
%\end{figure}

\subsection{Two-component fit to the bolometric LC: A support to the weak deflagration model with a bound remnant}\label{para}

\begin{figure}
  \begin{center}
   \includegraphics[width=120mm,clip,bb=0 0 360 270]{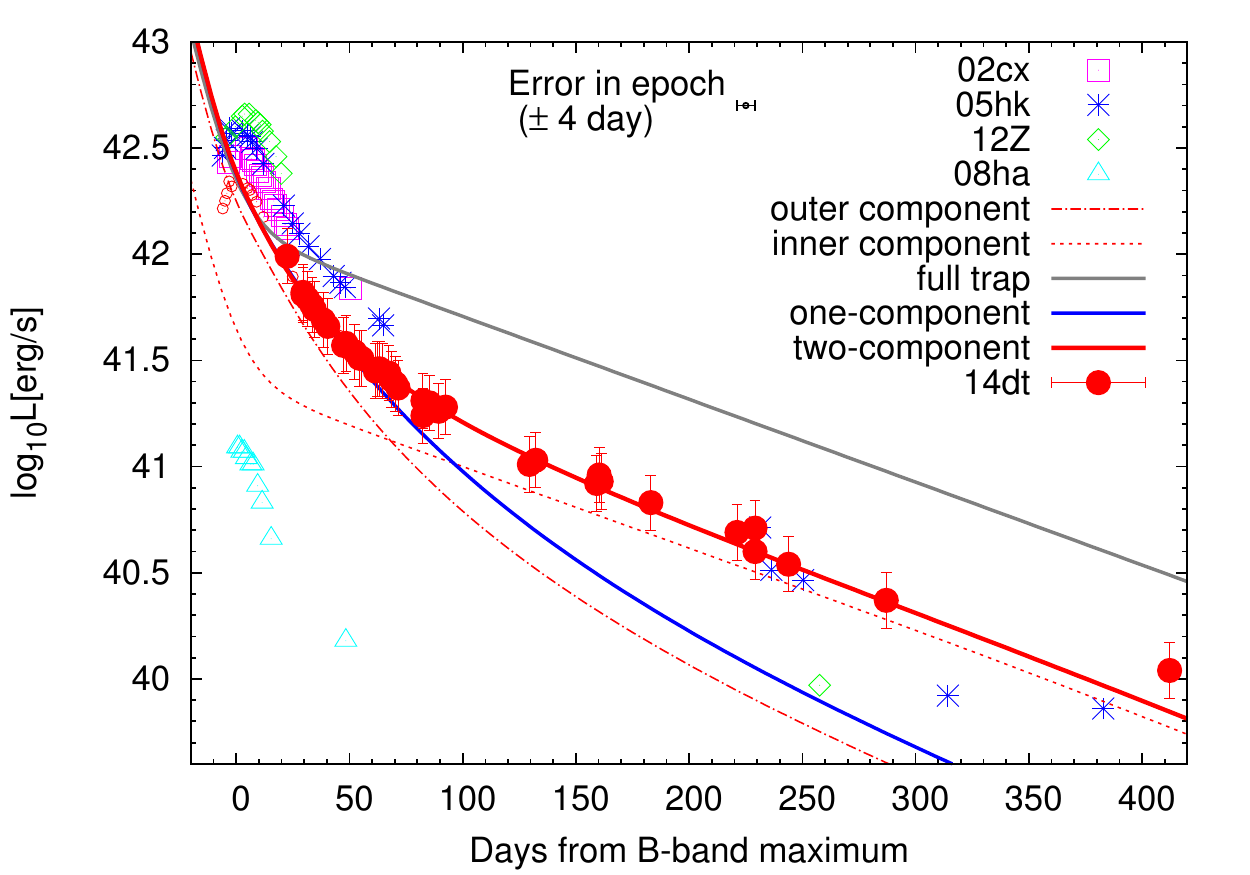}
  \end{center}
   \caption{The bolometric light curve of SN 2014dt.
This is the `combined' bolometric LC with the template LCs 
taken from SN 2005hk and $\Delta t = 0$ days (see \S 4.2 
and Appendix).
The red filled and open circles are the bolometric 
luminosities obtained from the sum of fluxes in the 
$BVRI$-bands and those from fitted template LCs in 
SN 2005hk, respectively.
At 232, 287 and 412 days, we calculated the bolometric 
luminosity from only $V$- and $R$-band fluxes under the 
assumption of the non-evolving SED shape at $\gtrsim 100$ 
days (see \S 3.5) because we did not have $B$- and $I$-band 
data.
For comparison, we plotted the LCs of SNe 2002cx (Li et al. 
2003), 2005hk (Sahu et al. 2008), 2008ha (Foley et al. 2009) 
and 2012Z (Yamanaka et al. 2015).
The gray, blue and red solid lines show the full-trap model, 
and one- and two-component LC models, respectively.
In these models, the total $^{56}$Ni mass is set to be 
$\sim$0.10 M$_{\odot}$.
The dashed and dotted lines indicate the outer and inner 
components, respectively, in the two-component model.
}\label{14dt_bol}
\end{figure}

We derived the bolometric luminosity of SN 2014dt assuming 
that the sum of fluxes in the $BVRI$-bands occupied about 
60 \%\ of the bolometric one (Wang et al. 2009).
The derived bolometric LC is shown in Figure \ref{14dt_bol}, 
which is one of the `combined LCs', with the template LCs 
taken from SN 2005hk and $\Delta t = 0$ days (see \S 3.3 
and Appendix).
In the early phase, SN 2014dt was $0.1$--$0.5$ dex fainter 
than SNe 2002cx, 2005hk, and 2012Z.
SN 2014dt showed a significantly slow decline after $\sim 60$ 
days, and then its luminosity became comparable with that of 
SN 2005hk as well as $\sim 0.5$ dex brighter than that of 
SN 2012Z at $230$--$250$ days.
With the peak luminosity ($L_{\rm bol, max}$) and rising 
time ($t_{r}$), we could estimate the mass of the synthesized 
$^{56}$Ni (Arnett 1982; Stritzinger \& Leibundgut 2005).
The rising part of the LC in SN 2014dt was not observed. 
SNe Iax show larger diversity, but it is pointed out that the 
rise time of SNe Iax has a correlation with the 
peak magnitude (Magee et al. 2016).
SN 2014dt is slightly fainter than SNe 2005hk and 2012Z, 
and it might be not a clear outlier like SN 2007qd (McClelland 
et al. 2010, Magee et al. 2016).
We assume that the rise time of SN 2014dt is roughly similar 
to the rise time of SNe 2005hk and 2012Z.
Adopting SN 2005hk and 2012Z as a template, we tested a 
range of rising times to estimate the mass of $^{56}$Ni. 
The definition of the rising time can differ in different 
literatures, so we decided to derive the rising time 
ourselves.
We fitted each of the multi-band LCs of SNe 2005hk and 
2012Z by a quadratic function, deriving rising times, and 
$t_{r} = 16.5$--$21.0$ and $11.8$--$12.8$ days for SNe 2005hk 
and 2012Z, respectively, which are consistent with those in 
previous papers.
Using $L_{\rm bol, max} =$ ($1.2$--$1.8$) 
$\times 10^{42}$ erg s$^{-1}$\footnotemark[14]\footnotetext[14]{
$L_{\rm bol, max} =$ ($0.3$--$1.7$) $\times 10^{42}$ erg s$^{-1}$ 
using the distance obtained the CO Tully-Fisher relation, 
$L_{\rm bol, max} =$ ($0.5$--$1.6$) $\times 10^{42}$ erg s$^{-1}$ 
using the distance obtained the H{\sc i} Tully-Fisher relation.}
and $t_{r} = 11.8$--$21.0$ days, we estimated the $^{56}$Ni mass as 
$0.04$--$0.10$M$_{\odot}$
\footnotemark[15]\footnotetext[15]{The $^{56}$Ni mass is 
$0.01$--$0.09$M$_{\odot}$ using the distance obtained 
the CO Tully-Fisher relation, $0.02$--$0.09$M$_{\odot}$ 
using the distance obtained the H{\sc i} Tully-Fisher 
relation.} for SN 2014dt.
It unfortunately has a large uncertainty, because the 
data around the maximum light are missing for SN 2014dt.

We also estimated the $^{56}$Ni mass from the late phase 
bolometric LC, which was subject to much less contamination 
by the error of the explosion epoch.
First, we estimated the $^{56}$Ni mass by a fit of a simple 
radioactive-decay LC model in the late phase (e.g., Maeda et al. 
2003),
\begin{eqnarray}
L_{\rm bol} &=& M(^{56}\mbox{Ni}) \left[ e^{(-t/8.8{\rm\ d})} \epsilon_{\gamma,\rm Ni} (1 - e^{-\tau}) \right. \nonumber \\
& & + \left. e^{(-t/113{\rm\ d})} \left\{ \epsilon_{\gamma,\rm Co} (1 - e^{-\tau}) + \epsilon_{e^{+}} \right\} \right] \mbox{, and}
\label{eq:one-l}
\end{eqnarray}
\begin{equation}
\tau \simeq 1000 \times \left[\frac{(M_{\rm ej}/M_{\odot})^{2}}{E_{51}}\right] (t \mbox{day})^{-2} \mbox{ ,}
\label{eq:tau}
\end{equation}
where $\epsilon_{\gamma, \rm Ni} = 3.9 \times 10^{10}$ erg s$^{-1}$ g$^{-1}$ 
is the energy deposition rate by $^{56}$Ni via $\gamma$-rays, 
$\epsilon_{\gamma, \rm Co} = 6.8 \times 10^{9}$ erg s$^{-1}$ g$^{-1}$ and 
$\epsilon_{e^{+}} = 2.4 \times 10^{8}$ erg s$^{-1}$ g$^{-1}$ 
are those by the $^{56}$Co decay via $\gamma$-rays and 
positron ejection, $M_{\rm ej}$ is the ejecta mass, and 
$E_{51}$ is the kinetic energy of $10^{51}$ erg.
However, it does not consistently explain 
the derived bolometric LC at both and early and late phases.
The slow tail in the obtained bolometric LC of SN 2014dt 
after $\sim 60$ days required nearly a full trapping 
of the deposited energy (that is, a larger $\tau$).
However, the full trapping of $\gamma$-rays 
with $\sim 0.10$ M$_{\odot}$ of $^{56}$Ni / $^{56}$Co (as 
derived above to fit the peak) should result in a much larger 
$L_{\rm bol}$ in the late phase (see Figure\ref{14dt_bol}).

Thus, we adopted a two-component LC model (Maeda et al. 2003) 
where $L_{\rm bol}$ is a sum of $L_{\rm bol, in}$ and 
$L_{\rm bol, out}$.
In this model, we have four independent parameters in total,
$M_{\rm in}(^{56}$Ni$), [(M_{\rm ej}/M_{\odot})^{2}/E_{51}]_{\rm in}$,
$M_{\rm out}(^{56}$Ni$)$ and $[(M_{\rm ej}/M_{\odot})^{2}/E_{51}]_{\rm out}$.
The former two parameters (subscripted by `in') correspond 
to an inner, larger $\tau$ component, and the latter two 
correspond to the outer, smaller $\tau$ one.
With this model, we successfully reproduced the LC which 
well traces the bolometric LC from near maximum to 
$\sim 250$ days as shown in Figure \ref{14dt_bol}.

The parameters obtained by the fit were 
$M_{\rm in}(^{56}$Ni$) =$ $0.015$--$0.025$ M$_{\odot}$,
$[(M_{\rm ej}/M_{\odot})^{2}/E_{51}]_{\rm in} \gtrsim 500$,
$M_{\rm out}(^{56}$Ni$) =$ $0.02$--$0.08$ M$_{\odot}$ and
$[(M_{\rm ej}/M_{\odot})^{2}/E_{51}]_{\rm out} \sim$ $0.8$--$3.0$.
Again, we noted that the errors were conservatively 
associated with a series of the allowed combined LCs for 
SN 2014dt (\S 3.3), which is also the case for the 
following analyses.
We could only estimate a rough lower-limit for 
$[(M_{\rm ej}/M_{\odot})^{2}/E_{51}]_{\rm in}$ so that it 
was large enough to fully trap the $\gamma$-rays even at 
$250$--$410$ days.
That is, the need for the high-density inner component was 
robust and insensitive to the parameters for fit.
At 410 days, the bolometric LC of SN 2014dt was slightly 
brighter than the two-component model LC.
Contamination by the galaxy background was small in this 
phase (see Foley et al. 2014). 
If this excess was real, one possible explanation is the 
contribution from a scattering component of a presumed CSM 
echo, corresponding to the MIR excess (\S 4.1).

The existence of the large $\tau_{\rm in}$ component 
suggests that there is an inner component that keeps a 
high density even in the late phase, regardless of the 
origin (see \S 4.1).
This is consistent with the prediction of the weak 
deflagration model with a bound remnant (e.g., 
Fink et al. 2014).

\subsection{Properties of Ejecta}

In the two-component model, the inner, dense component may 
be connected with the existence of a bound white dwarf, 
and the outer, less dense one may correspond to the SN ejecta.
Assuming that the outer component
($[(M_{\rm ej}/M_{\odot})^{2}/E_{51}]_{\rm out} \sim$ 
$0.8$--$3.0$) was the main ejecta, we could 
estimate $M_{\rm ej}$ and $E_{\rm k}$ of the ejecta in SN 
2014dt by applying the scaling laws,
\begin{equation}
t_{\rm d} \propto \kappa^{1/2} \ M_{\rm ej}^{3/4} \ E_{\rm k}^{-1/4} \mbox{, and}
\label {eq:difftime}
\end{equation}
\begin{equation}
v \propto E_{\rm k}^{1/2} \ M_{\rm ej}^{-1/2} \mbox{,}
\label {eq:velocity}
\end{equation}
calibrated with the well-studied SNe Ia, where $t_{\rm d}$ 
is the diffusion timescale, $\kappa$ is the absorption 
coefficient for optical photons, and $v$ is the typical 
expansion velocity of the ejecta.
Here, we adopted the parameters derived for SN Ia 2011fe 
(Pereira et al. 2013) for the normalization.
From Equations (\ref{eq:difftime}) and (\ref{eq:velocity}),
we obtained the following equations of the ratios in 
$M_{\rm ej}$ and $E_{\rm k}$,
\begin{equation}
\frac{M_{\rm ej, 14dt}}{M_{\rm ej, 11fe}} = \left( \frac{t_{\rm d, 14dt}}{t_{\rm d, 11fe}} \right)^{2} \times \frac{v_{\rm 14dt}}{v_{\rm 11fe}} \times \frac{\kappa_{\rm 11fe}}{\kappa_{\rm 14dt}} \mbox{, and}
\label {eq:ej_mass}
\end{equation}
\begin{equation}
\frac{E_{\rm k, 14dt}}{E_{\rm k, 11fe}} = \left( \frac{t_{\rm d, 14dt}}{t_{\rm d,11fe}} \right)^{2} \times \left( \frac{v_{\rm 14dt}}{v_{\rm 11fe}} \right)^{3} \times \frac{\kappa_{\rm 11fe}}{\kappa_{\rm 14dt}} \mbox{.}
\label {eq:kin_energy}
\end{equation}

The expansion velocity of a SN Ia is usually estimated 
from the blueshift of the Si {\sc ii} $\lambda$6355 
absorption line at $\sim 0$ day (e.g., 
$v_{\rm 11fe} \simeq 10500$ km s$^{-1}$; 
Pereira et al. 2013).
The Si {\sc ii} $\lambda$6355 absorption line of SN Ia is 
strong at $\sim 0$ day and is little contaminated by 
other absorption lines (e.g, SN 2011fe; Parrent et al. 2012). 
However, we have no spectrum of SN 2014dt around the maximum.
In addition, it is often difficult to identify the 
Si {\sc ii} $\lambda$6355 line in the spectra of SNe Iax 
at 10 days and later.
Thus, we estimated the Si {\sc ii} $\lambda$6355 line 
velocity, assuming that the ratio between the velocity of 
Si {\sc ii} $\lambda$6355 at $\sim 0$ day and that of 
Fe {\sc ii} lines velocity at $\sim 20$ days was similar 
among SNe Iax samples (i.e., SNe 2005hk, 2012Z and 2014dt).
In SNe Iax, the Fe {\sc ii} $\lambda$6149 and $\lambda$6247 
lines were not much blended (see \S3.2).
In this way, we estimated $v_{\rm 14dt} = 3500$--$4200$ 
km s$^{-1}$.

The diffusion timescale, $t_{\rm d}$, approximately corresponds
to the width of the peak in the bolometric LC.
We derived the ratio of $t_{\rm d, 14dt}/t_{\rm d, 11fe}$ 
as $1.0$--$1.6$, by comparing the durations for the two SNe 
from the maximum to the epoch when the magnitude decreased 
by 0.5 dex. 

For $\kappa$, it would not be safe to simply assume that 
the opacity of SN 2014dt is the same as that of SN 2011fe, 
because the properties of SN ejecta would be considerably 
different between SNe Iax and normal SNe Ia (i.e., 
$\kappa_{\rm 14dt} \neq \kappa_{\rm 11fe}$).
Thus, we kept it as a free parameter and added another 
observational constraint using the post-maximum LC, when 
the LC is still dominated by the ejecta (that is, the 
outer component).
With $E_{\rm k, 11fe} = 1.2 \times 10^{51}$ erg (Pereira 
et al. 2013) and $M_{\rm ej, 11fe} = 1.4$ M$_{\odot}$, we 
obtained $E_{\rm k, 14dt} =$ ($0.07$--$0.42$) $\times 10^{50}$ 
erg, $M_{\rm ej, 14dt} = 0.08$--$0.35$ M$_{\odot}$ and
$\kappa_{\rm 14dt} / \kappa_{\rm 11fe} = 1.5$--$11.6$.
The relatively large uncertainties reflect the large 
errors in the estimated explosion epoch (see \S 3.4).
Similarly, we obtained 
$\kappa_{\rm 05hk} / \kappa_{\rm 11fe} = 2.2$--$4.3$
(see \S 4.5 for details).
Interestingly, the opacity of the SNe Iax was found to be 
larger than that of normal SNe Ia.
Foley et al. (2013) found the ejecta mass of SNe Iax is 
systematically small based on the LC analysis\footnotemark[16]
\footnotetext[16]{In previous paper, some authors pointed 
out that the ejecta mass of SNe Iax is consistent with a 
Chandrasekhar-mass (e.g., Sahu et al. 2008, Stritzinger et al. 2015).}.
The contribution of the iron-group elements to 
the opacity could differ from those in the ejecta of 
normal SNe Ia.
This large ratio $\kappa_{\rm Iax} / \kappa_{\rm 11fe}$ 
indicates that a difference exists.
However, for a more realistic treatment, one has to take into 
account the different physical properties (e.g., ionization) 
in SNe Iax and normal SNe Ia, which is beyond the scope of 
this study.

\subsection{Explosion Mechanism}

We discuss the explosion mechanism of SN 2014dt based on 
the explosion parameters derived above.
Since the derived explosion parameters have a large 
uncertainty, the discussion here should be regarded as 
being qualitative.
For SNe Iax, several different explosion models have been 
suggested (e.g., the weak deflagration model with or 
without a bound remnant, the pulsating delayed detonation, 
and the fallback CC models).
For the weak deflagration model without a bound remnant 
(Nomoto et al. 1976; Branch et al. 2004; Jha et al. 2006; 
Sahu et al. 2008), our analysis suggests that this model 
clould not explain the slow decline LC in the late phase 
(see \S 4.2).
As previously mentioned, the observational properties of 
SNe Iax are explained by the pulsational delayed 
detonation model (PDD model; H\"oeflich et al. 1995, H\"oeflich 
\& Khokhlov 1996) (e.g., SN 2012Z; Stritzinger et al. 2015). 
In the PDD model, the detonation is triggered after the 
deflagration phase. 
This creates a high-density inner ejecta, and thus it may explain 
the slow evolution and low velocities seen in the late time 
spectra of SN 2014dt.

In Table 6, we show the characteristic explosion parameters 
predicted by the weak deflagration model with a bound remnant 
(Fink et al. 2014) and by the fallback CC SN model (Moriya 
et al. 2010), which are apparently consistent with those 
derived for SN 2014dt.
In the fallback CC model, some progenitor models (a He star 
with initial mass of $40$ M$_{\odot}$, CO stars with 
initial masses of $25$ and $40$ M$_{\odot}$) seem to 
explain the observations.
However, Foley et al. (2015) gave the upper-limit of the 
brightness of the progenitor from the pre-explosion images 
taken by the HST and suggested that an existence of a 
massive star in the progenitor system is unlikely.
Thus, we consider that the weak deflagration model with a 
bound remnant is most plausible for SN 2014dt.
However, the explosion parameters of SN 2014dt have large 
uncertainties due to a lack of pre-maximum data.
Thus, we do not argue strongly that the derived parameters 
provide a strong evidence of the weak deflagration scenario.
Rather, we list this analysis as a supporting evidence to 
the arguments in the late phase behaviors, which led us to 
suggest the weak deflagration model for SNe 2014dt and 2005hk.
The slow evolution in the SED of SN 2014dt in the same phase 
(\S 4.1) seems consistent with this model.

Our late time data revealed the nearly unchanged 
SED and full $\gamma$-ray trapping, which are difficult to explain 
by the expanding SN ejecta with decreasing density.
It is thus tempting to connect these properties to the weak 
deflagration model, in which the bound remnant and its atmosphere  
may serve as a high-density ``stationary'' source of the energy input.
If the PDD model would also create a slow, high-density Fe-rich 
ejecta which keeps opaque for 400 days, this could also be a viable 
model, while this may require fine-tuning in the model. 
We suggest the weak deflagration as a straightforward interpretation, 
but distinguishing these two scenarios will require further study both 
in theory and model. 
%For example, the weak deflagration model by Fink et al. (2014) leads 
%to a rapid decline in the post-maximum phase, without including the 
%effect of the central remnant. 
%The low ejecta is not enough to trap the $\gamma$-ray, the model could 
%not reproduce the LCs in redder bands.
%In addition, the model predicts a non-layered ejecta structure due to 
%the mixing.
%Barna et al. (2017) suggested the ejecta separate into a well-mixed 
%region and a stratified region from the spectral model fitting.
%SN 2012Z shows the flat velocity evolution of the intermediate mass 
%elements, it indicates the ejecta has the layered stracture (Stritzinger 
%et al. 2015).
%These indicate the ejecta of some bright SNe Iax have the layered stracture.
%Investigation of the effects of the bound remnant in the observables 
%is thus strongly encouraged. 
%In the observational side, the two scenarios may predict the difference 
%in the formation of the low velocity permitted lines in the late phase. 
Indeed, there are two possible drawbacks so far raised 
for the weak deflagration model. 
(1) The weak deflagration model by Fink et al. (2014) leads to a rapid 
decline in the post-maximum phase. 
However, the model does not include the effect of the central remnant, 
and thus the investigation of the effects of the bound remnant 
in the LC is strongly encouraged. 
(2) The weak deflagration model inevitably leads to a well-mixed, 
non-stratified structure in the ejecta, while the layered structure has 
been inferred at least for some SNe Iax from the velocity evolution 
(SN 2012Z; Stritzinger et al. 2015) and through spectral modeling 
(SN 2011ay; Barna et al. 2017). 
Indeed, this feature has been suggested to be consistent with the PDD 
model rather than the weak deflagration model (Stritzinger et al. 2015). 
To further add possible diagnostics to differentiate the two models, 
we encourage to investigate the evolution of line velocities in the 
late phase. 
The two scenarios may predict the difference in the formation of the 
low velocity permitted lines in the late phase.
The weak deflagration model could produce them either in the remnant 
WD and the slow expanding ejecta, while only the latter is possible 
in the PDD.
As these two mechanism will lead to different evolution in the line 
velocities, a high-resolution spectroscopy in the late phase 
may solve this issue.

\begin{longtable}{lccc}
\caption{Comparison of explosion parameters with candidate models}
\hline
Model & Explosion energy ($10^{50}$erg) & Ejecta mass ($M_{\odot}$) & $^{56}$Ni mass ($M_{\odot}$)\\
\hline
\endhead
\hline
\endfoot
\hline
\multicolumn{4}{l}{\hbox to 0pt{\parbox{180mm}{\footnotesize
   \footnotemark[17] With a bound remnant. N1def and N3def models for Fink et al. (2014)
   \par\noindent
   \footnotemark[18] He star with initial mass of 40 M$_{\odot}$,
CO stars with initial masses of $25, 40$ M$_{\odot}$ for Moriya et al. (2010)
   \par\noindent
   \footnotemark[19] This study. $^{56}$Ni mass of SN 2005hk is referred to Sahu et al. (2008).
}\hss}}
\endlastfoot
  Weak deflagration \footnotemark[17] & $0.149$--$0.439$ & $0.0843$--$0.195$ & $0.0345$--$0.0730$ \\
  Core collapse (40He) \footnotemark[18] & $0.12$--$0.33$  & $0.10$--$0.23$ & - \\
  Core collapse (25CO) \footnotemark[18] & $0.073$--$0.33$ & $0.056$--$0.24$ & - \\
  Core collapse (40CO) \footnotemark[18] & $0.038$--$0.39$ & $0.029$--$0.25$ & - \\ \hline
  SN 2014dt \footnotemark[19] & $0.07$--$0.42$ & $0.08$--$0.35$ & $0.04$--$0.10$ \\
  SN 2005hk \footnotemark[19] & $0.42$--$0.88$ & $0.21$--$0.42$ & $0.17 \pm 0.02$ \\ \hline
\end{longtable}

\subsection{Comparison with another SN Iax 2005hk}

It is interesting seen in discuss the diversities 
in the observed properties of SNe Iax 2005hk, 2012Z and 2014dt 
from a viewpoint of the prediction of the weak deflagration model.
As shown in \S 3, the diversities in photometric and 
spectroscopic data are likely more significant in 
the late phase than in the early phase.
In this subsection we apply the modeling procedure adopted 
in the previous sections to another SN Iax 2005hk, to 
investigate whether SN 2005hk can be explained within the 
same context of the weak deflagration model.

\begin{figure}
  \begin{center}
   \includegraphics[width=120mm,clip,bb=0 0 360 270]{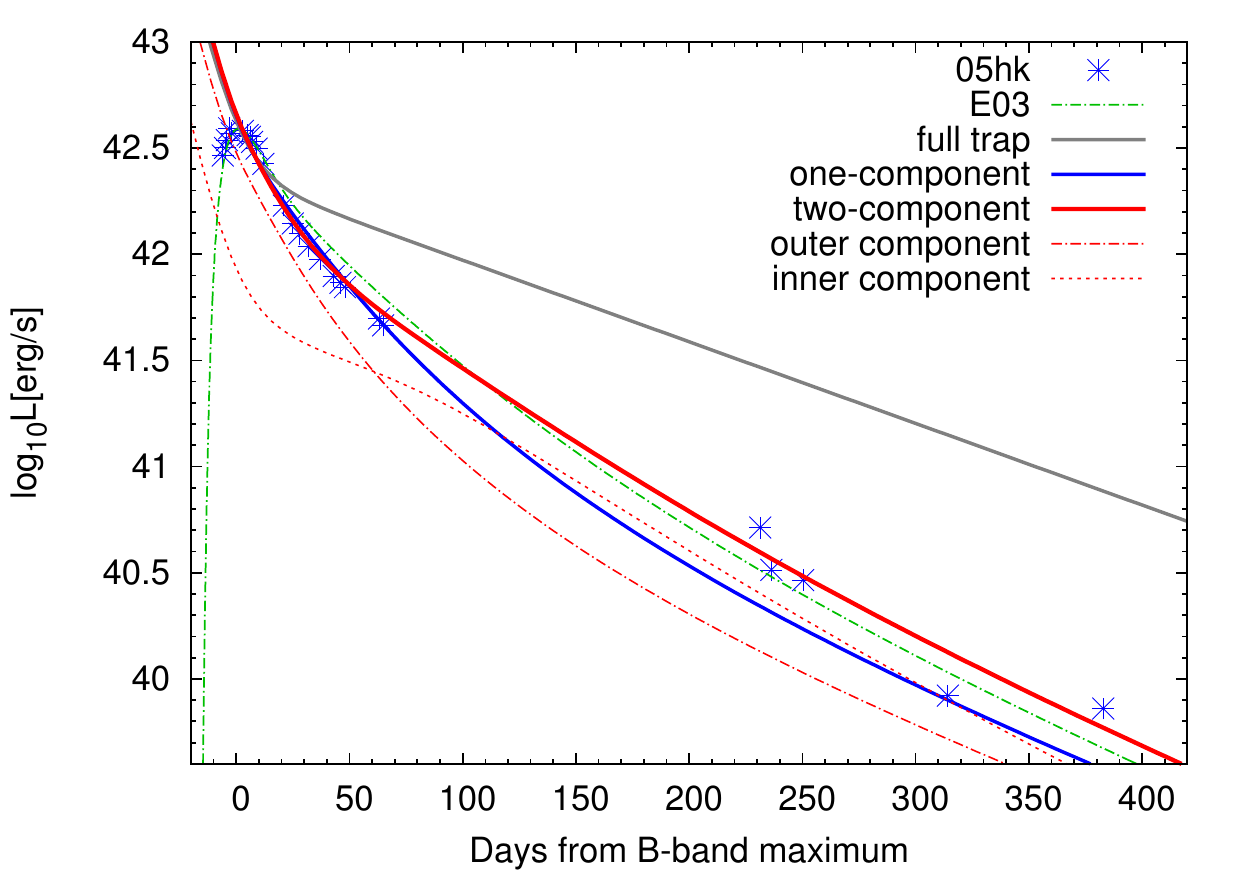}
  \end{center}
   \caption{The bolometric light curve of SN 2005hk and 
corresponding LC models. 
The lines are plotted in the same manner as 
in Figure \ref{14dt_bol}.
The total $^{56}$Ni mass is 0.18 M$_{\odot}$.
We also plotted the LC of the E03 model 
(weak deflagration model without a bound remnant; 
Sahu et al. 2008) by a green dashed line (see \S 
4.5), which also well represents the entire LC from 
the early through the late phases.
}\label{05hk_bol}
\end{figure}

For SN 2005hk, Sahu et al. (2008) suggested the weak 
deflagration model without a bound remnant, which could 
reproduce the LC. 
The LC model `E03' in Figure 11 is the LC 
presented by Sahu et al. (2008), where the explosion 
parameters are $E_{\rm k, 05hk} = 3.0 \times 10^{50}$ erg 
and $M_{\rm ej, 05hk} = 1.4$ M$_{\odot}$.
In the sequence of the weak deflagration models by Fink 
et al. (2014), it is expected that such a small $E_{\rm k}$ 
model would leave a bound remnant. 
The critical value is $\sim $($6.0$--$7.0$) $\times 10^{50}$ 
erg (Fink et al. 2014).
However, considering the uncertainty in the model, it is 
difficult to conclude that kinetic energy as small as 
$3.0 \times 10^{50}$ erg should be associated with a bound 
remnant.

Alternatively, a two-component model can also explain the 
LC of SN 2005hk, which mimics the presence of a bound 
remnant (Figure \ref{05hk_bol}).
The parameters of the two-component model are 
$M_{\rm in}(^{56}$Ni$) \sim 0.04$ M$_{\odot}$,
$[(M_{\rm ej}/M_{\odot})^{2}/E_{51}]_{\rm in} \sim 30$,
$M_{\rm out}(^{56}$Ni$) \sim 0.14$ M$_{\odot}$ and
$[(M_{\rm ej}/M_{\odot})^{2}/E_{51}]_{\rm out} \sim 1.0$--$2.0$.
SN 2005hk has a smaller fraction of the $^{56}$Ni mass 
in the inner component than SN 2014dt.
In the same way, we estimated the explosion parameters for 
SN 2005hk as $E_{\rm k, 05hk} = $($0.42$--$0.88$) $\times 10^{50}$ 
erg and $M_{\rm ej, 05hk} = 0.21$--$0.42$ M$_{\odot}$ within 
the context of the two-component model.
The derived $[(M_{\rm ej}/M_{\odot})^{2}/E_{51}]_{\rm in}$ 
indicates that the average density in the 
inner component of SN 2014dt may be larger 
than that of SN 2005hk, consistent with indications from the 
late phase spectra (\S 3.2).
The $E_{\rm k, 05hk}$ and $M_{\rm ej, 05hk}$ are 
consistent with the weak deflagration model with a bound 
remnant, whose mass is smaller than in SN 2014dt (Fink 
et al. 2014).

In summary, we conclude that SN 2005hk can be explained 
by the weak deflagration model as well. 
It is unclear whether SN 2005hk left a bound remnant.
If a bound remnant was left, the mass of the remnant 
would be smaller than in SN 2014dt.
  
\section{Conclusions}

We presented long-term optical and NIR observations of 
SN 2014dt up to 410 days after the maximum light.
The data for SNe Iax in the NIR bands are rare, 
especially in the late phase.
We could obtain the NIR evolution for SN 2014dt until 
the late phase.
The LC showed a considerably slow decline in the late phase; 
the decline rate was the smallest among well-studied SNe 
Iax samples.
From the evolution of the SED, SN 2014dt did not show 
significant change in the BB temperature.
The spectral features were also slow in SN 2014dt in 
the late phase.
A bound remnant left after the explosion gives a good 
account of these observational properties.

By scaling the explosion parameters of normal SN Ia 2011fe,
we suggest that the mass of synthesized $^{56}$Ni was 
$0.04$--$0.10$ M$_{\odot}$, the kinetic energy was 
($0.07$--$0.42$) $\times 10^{50}$ erg, and the ejecta mass 
was $M_{\rm ej} = 0.08$--$0.35$ M$_{\odot}$.
These values are consistent with the prediction 
of the weak deflagration model with a bound remnant.
However, the uncertainties in the derived explosion 
parameters for SN 2014dt were large, reflecting our 
conservative error estimation associated with 
a possible range of the missed pre-maximum LC behavior.

In summary, the overall properties derived from the 
long-term observations of SN 2014dt could be explained 
with the weak deflagration model, with leaves a bound 
remnant after the explosion.
Indeed, by applying the same model to SN 2005hk, we found 
that SN 2005hk could also be explained by the weak deflagration 
model with an explosion energy larger than for SN 2014dt.
With the larger kinetic energy, it might not leave a bound 
remnant.
The present work demonstrates the power and importance of late 
phase observations to uncover the origin of SNe Iax.
To further test the explosion models, we need to obtain 
a larger sample of well-observed SNe Iax in the future.

While the weak deflagration model with the bound 
remnant provides straightforward interpretation to our data, 
further study is required to robustly identify the explosion 
mechanism. 
For example, the PDD model may also account for main observational 
features including those derived from our data set. 
For example, further study on the predicted observable based on 
large model grids both for the deflagration and PDD scenarios will 
help to resolve the issue, including the synthetic light curves for 
the deflagration model including the contribution from the bound 
remnant. 
Observationally, a key would be high spectral-resolution observations 
in the late phase to robustly measure the velocities of the low 
velocity permitted lines and their evolution, which would be used to 
identify the origin of the late phase slow evolution in the SED 
either as being caused by the low velocity ejecta or the WD wind.

\section*{Acknowledgements}
We thank K. Aoki, Y. Utsumi and members of observation team 
at OKU for their helps with the observations.
We thank the staff at the Subaru Telescope for their 
excellent support of the observations under S15A-078 and 
S15B-055 (PI: K. Maeda).
The authors also thank D. K. Sahu and M. Tanaka for insightful 
comments and kindly providing the bolometric LC data of SN 2005hk 
and the result of their model calculation.
The spectral data of comparison SNe are downloaded from 
SUSPECT\footnotemark[20]\footnotetext[20]{http://www.nhn.ou.edu/\~{}suspect/}(Richardson et al. 2001) and 
WISeREP\footnotemark[21]\footnotetext[21]{http://wiserep.weizmann.ac.il/}(Yaron \& Gal-Yam 2012) databases 
and the UC Berkeley Filippenko Group's Supernova Database\footnotemark[22]\footnotetext[22]{http://heracles.astro.berkeley.edu/sndb/}, 
the photmetry data are downloaded from ``The Open Supernova Catalog'' database\footnotemark[23]\footnotetext[23]{https://sne.space} 
(Guillochon et al. 2017).
This research has made use of the NASA/IPAC Extragalactic 
Database (NED) which is operated by the Jet Propulsion Laboratory, 
California Institute of Technology, under contract with the National 
Aeronautics and Space Administration.
This work is supported by the Grant-in-Aid for Scientific Research
from JSPS (JP23244030, JP26287031, JP26800100, JP26400222, JP16H02168, 
JP17K05382, JP17H02864, JP18H04585, JP18H05223), and by Optical and 
NIR Astronomy Inter-University Cooperation Program, OISTER, the Hirao 
Taro Foundation of the Konan University Association for Academic 
Research, and by WPI Initiative, from the Ministry of 
Education, Culture, Sports, Science and Technology in Japan.

\appendix

\section{Estimate of the Pre-maximum LC Behavior}\label{sec:lc_matching}

Here we describe how to constrain the maximum brightness and 
the maximum date of SN 2014dt. 
We compared its early multi-band LCs with 
those of well-observed SNe Iax 2002cx, 2005hk, and 2012Z for 
which the spectra are similar to those of SN 2014dt.

We adopted the multi-band LCs of SNe 2002cx, 2005hk, and 
2012Z as templates because those LCs cover both their own 
maximum phases and the phase when the data of SN 2014dt exist.
Moreover, there are spectral resemblances and apparent 
similarities in the decaying part of the multi-band LCs 
(see \S 3.1 -- 3.3). 
Our initial guess of the $B$-band maximum date of SN 2014dt 
was MJD 56950.8 (20.8 October 2014) in terms of the similarity 
in the spectral phase (\S 3.3). 
For a LC of each SN in a given bandpass, we allowed a shift 
in the time axis $\Delta t$ and in the magnitude axis 
$\Delta M$. 
Here, $\Delta t = 0$ corresponds to the initial guess, and 
we allowed the range of $\Delta t = \pm 10$ days with an 
interval of 2 days.
In this procedure, the relatively densely-sampled LC data 
of SN 2014dt were interpolated to provide the magnitude at 
the (assumed) same epochs with the data points available 
for the reference/template SNe Iax. 
For each set of $\Delta t$ and $\Delta M$, we computed a 
residual between the SN 2014dt and the hypothesized template 
LC between $\sim 20$ (the discovery date) and 60 days after 
the discovery of SN 2014dt. 
This residual, $\bar{M}$ is defined as follows; 
\begin{equation}
\bar{M} = \sqrt[]{\mathstrut \Sigma \frac{{(m_{Iax} - m_{14dt})}^{2}}{N}} \mbox{,}
\label {eq:chi}
\end{equation}
where $m_{Iax}$ is the $BVRI$-band magnitude of the reference 
SN Iax, $m_{14dt}$ is that of SN 2014dt, and $N$ is the number 
of data points of SN 2014dt that overlap with the template LCs.
We first obtained the `canonical magnitude offset' in the 
magnitude axis, where $\bar{M}$ takes the minimum value 
($\Delta M_{0}$) for given $\Delta t$. 
To further evaluate the uncertainty in the magnitude axis 
in the fit, we varied the magnitude axis by $\Delta m$ 
against $\Delta M_{0}$ within the range of $\pm 0.3$ mag 
with an interval of $0.1$ mag, and then adopted 
$\Delta M = \Delta M_{0} + \Delta m$.
These procedures provided the distribution of $\bar{M}$ as 
an indicator of the similarity of the LCs among SN 2014dt 
and the reference SNe in each band separately ($BVRI$-band), 
as a function of $\Delta t$ and $\Delta M$. 
The set of $\Delta t$ and $\Delta M$ that provided the 
minimum $\bar{M} \equiv \bar{M_0}$ was adopted as our 
tentative best-fit model for each band. 
To evaluate their uncertainties and avoid possible local 
minima, we searched the area in the $\Delta t$--$\Delta M$ 
plane ($|\Delta t| \leq 10$ days and $|\Delta M| \leq 0.3$ 
mag) where $\bar{M} < 2 \times \bar{M_0}$ holds. 
The ensemble of the set of $\Delta t$ and $\Delta M$ 
satisfying above condition was adopted as acceptable ones 
and we estimated the uncertainties from their ranges. 
The choice of this criterion is somewhat arbitrary, but 
provides a reasonable estimate as judged by visual 
inspection of the fitting results.

Performing the same procedure independently for $BVRI$-band 
LCs, we determined the allowed range of $\Delta t$ and 
$\Delta M$ so that the fits in all bands were mutually 
consistent.
Indeed, we found that the multi-band LCs of SN 2002cx did 
not provide a consistent solution in each band (see below), 
and thus we concluded that the LCs of SN 2002cx should be 
omitted as templates. 
As a result, we have a series of `combined' LCs for SN 2014dt, 
in which the pre-discovery part was replaced by the LC of 
SN 2005hk or SN 2012Z, taking into account the uncertainty 
in $\Delta t$ and $\Delta M$.

The derived range of acceptance for $\Delta t$ is 
$\Delta t = -2$ $\sim$ $+4$ and $-4$ $\sim$ $0$, 
for SNe 2005hk and 2012Z, respectively, as templates. 
From these acceptance ranges, we assumed an uncertainty 
of $\pm 4$ days in the maximum epoch throughout this paper. 
Thus, we adopted the epoch of the $B$-band maximum and 
its error as MJD $56950.8 \pm 4.0$. 
Then, we have a series of `combined' LCs in which the LCs 
of SN 2014dt in the pre-discovery epoch were reconstructed 
using the LCs of a reference SN (either SN 2005hk or 2012Z), 
taking into account the conservative uncertainty in $\Delta t$ 
and $\Delta M$.
Throughout this paper, in cases where the pre-maximum LC 
data are used for the relevant analyses (e.g., the maximum 
magnitudes, explosion parameters), our estimation includes 
the error associated with this uncertainty (also including 
the difference in results between the two reference SNe).

To demonstrate the fitting result, and especially how it 
was dependent on the template SNe, we showed the best-fit 
LCs for each template SN (including SN 2002cx) in Figure 12. 
In the $B$- and $V$-bands, all of the reference SNe provided 
an acceptable fit for SN 2014dt. 
However, in the $R$- and $I$-bands, the LCs of SN 2002cx 
showed significantly slower decline than those of SN 2014dt 
and never provided a good fit, as was clear even by visual 
inspection.

\begin{figure}
  \begin{center}
   \includegraphics[width=120mm,clip,bb=0 0 360 270]{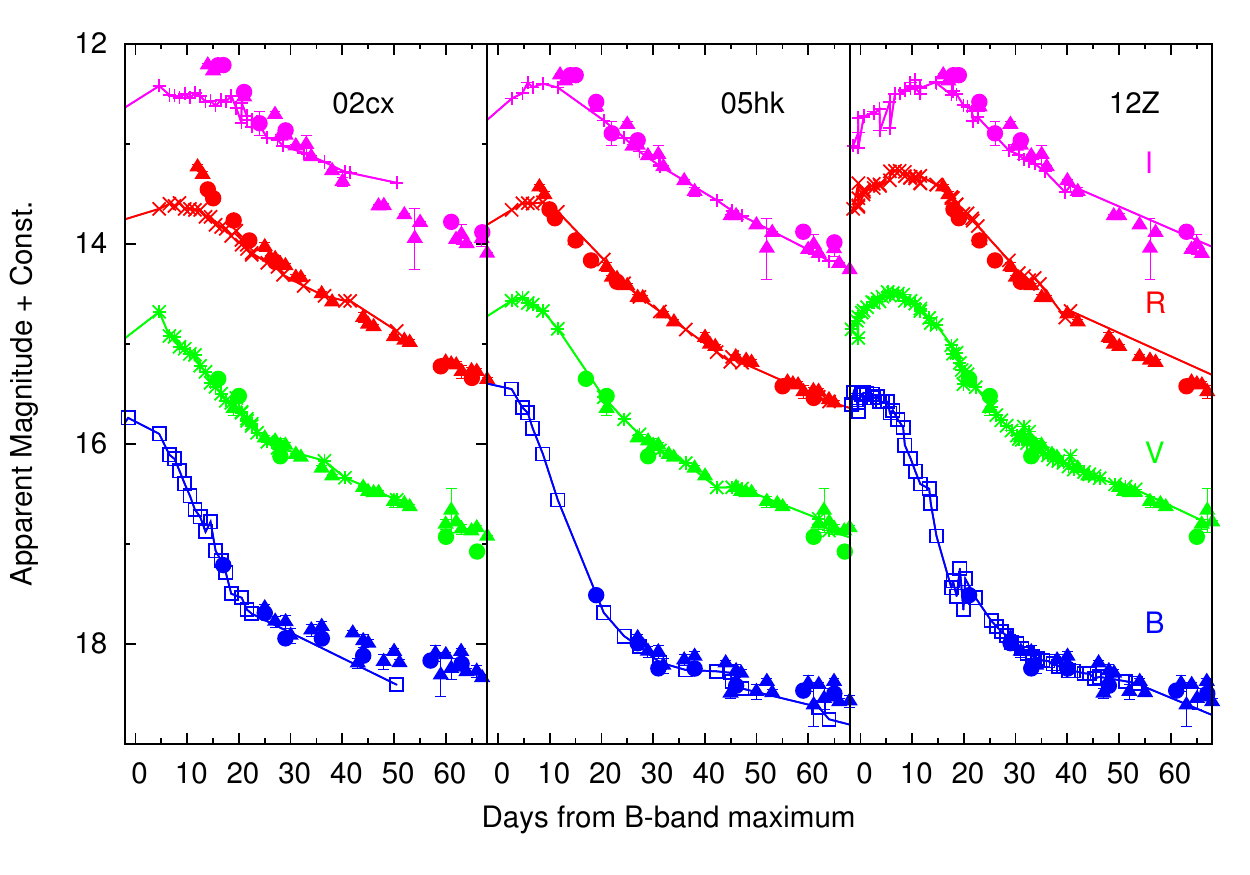}
  \end{center}
   \caption{The best-fit LCs of SN 2014dt in the $BVRI$-band.
These were derived in order to construct `combined' LCs in 
the period before the discovery (i.e., around the maximum 
of SN 2014dt).
We adopt the LCs of SNe 2002cx (left panel), 2005hk (middle) 
and 2012Z (right) as templates (see Appendix for details).
For SN 2014dt, the symbols are plotted in the same manner 
as in Figure \ref{14dt_lc}.
We show the fitted reference LCs by crosses connected 
by solid lines.}\label{bestfit}
\end{figure}

{}

\end{document}